# Lubricated friction of textured soft substrates

Yunhu Peng, Christopher M. Serfass, Catherine N. Hill, Lilian C. Hsiao*

Department of Chemical and Biomolecular Engineering, North Carolina State University, Raleigh, NC 27695, USA

*Corresponding author.

Email: lilian_hsiao@ncsu.edu

**The understanding of sliding friction for wet, patterned surfaces from first principles is challenging. While emerging applications have sought design principles from biology, a general framework is lacking because soft interfaces experience a multiphysics coupling between solid deformation and fluid dissipation. We investigate the elastohydrodynamic sliding of >50 patterned sliding pairs comprising elastomers, thermosets, and hydrogels, and discover that texturing induces a critical transition in the macroscopic friction coefficient. This critical friction scales universally, without any fitting parameters, with the reduced elastic modulus and the pattern geometry. To capture the frictional dissipation, we separate the flow curve into two regimes and account for the contributions of shear and normal forces applied by the fluid on the patterns. Our model combines Reynolds' equations and elastic deformation to provide physical insights that allow engineering of the elastohydrodynamic friction in a class of soft tribopairs using pattern geometry, material elasticity, and fluid properties.**

Patterned surfaces are ubiquitous in many applications. Fingerprint textures are thought to promote self-renewal and enhance haptic perceptions[1]. Tires and journal bearings are designed with textures to alter friction[2,3]. Arrays of micropillars infused with perfluorinated liquids, designed to mimic the interiors of pitcher plants, have generated extremely slippery behavior[4]. These examples suggest that a key function of surface textures is to control the dissipation of shear and normal stresses in the presence of a thin layer of lubricant. The liquid film generates fluid pressures that lift yet deform the solid surfaces, in a regime of tribology (the study of friction) known as elastohydrodynamic lubrication (EHL)[5,6]. Low-Reynolds number EHL flows are omnipresent in systems ranging from biological tissues to elastomers and ultrasoft hydrogels[7,8]. They are important in influencing the power consumption of machines and the physical sensations we feel on a daily basis.

Experimental studies show rough and patterned surfaces exhibit EHL tribology that is different from that of flat surfaces[9,10]. Although numerical simulations are able to closely reproduce the complex conditions within sliding tribopairs[11], these methods tend to be computationally expensive and provide little mechanistic insights into experimental systems. Our work is designed to provide a practical scaling framework in which friction is obtained from material elasticity and pattern geometry without the need for simulations. We observe that patterns aligned orthogonally to the flow velocity prevent free drainage of the lubricant and generate lift. To understand these effects, we turn to Reynolds' lubrication theory, which predicts that the shear force $F_S$ scales as $U/H$ and the normal force $F_N$ scales as $U/H^2$ (Fig. S1, Supplementary Information). Here, $U$ is the relative sliding speed between two tribopairs and $H$ is the lubricant film thickness. The ratio of $F_S$ and $F_N$ characterizes the bulk friction coefficient ($\mu$) of a tribopair, where miniscule changes in $H$ generate significant changes in $\mu$. Mesh-based simulations are often used to obtain the values of $H$ for flat tribopairs[12], and these methods have provided a satisfactory prediction of $\mu$ that matches experimental data in the EHL regime[11]. Nevertheless, the application of lubrication theory alone is insufficient for patterned tribopairs: the fluid should compress the raised textures more than the bulk substrate as sliding speed increases, resulting in non-monotonic changes in surface morphology and film thickness. Despite the importance of texture, the *a priori* prediction of $\mu$ for soft mesopatterned substrates in the EHL regime has not yet been demonstrated.

Fig. 1a shows how pattern compression gives rise to a completely different frictional behavior in the EHL regime. This so-called Stribeck curve is used to characterize the steady-state friction of tribopairs as a function of the Sommerfeld number, $S$. The dimensionless number $S = \eta U(4R^2/H)/F_N$ represents the relative contribution of fluid lubrication with respect to other frictional mechanisms, where $\eta$ is the average lubricant viscosity and $R$ is the radius of the contact area. EHL dominates the frictional dissipation at $S \gg 1$ while boundary friction dominates at $S \ll 1$. As $S$ increases, the steady-state $\mu$ of tribopairs transitions from a static value generated by solid-solid asperity contact to a monotonically increasing value in the EHL regime generated by viscous drag (Fig. 1a, Inset). Although birefringence and interferometry are used to measure $H$[6,13-15], which sets $S$ and therefore $\mu$, such techniques require complex instrumental setups where slight deviations can result in large uncertainties. These uncertainties in $H$ due to elastohydrodynamic deformation within the contact area are typically validated using simulations. However, finite element algorithms are resource-intensive for complex geometries[16], may not fully capture the effect of wall slip[9,13], and more importantly do not provide an understanding of the underlying physics responsible for the frictional properties of soft patterned materials.

We use a stress-controlled triborheometer to measure the EHL friction of 54 textured surfaces spanning four types of materials (silicones, hydrogels, polyesters, and mercaptoesters) with different elasticities and wettabilities (Fig. 1b). Patterned substrates are fabricated using standard lithography (Methods), then mounted onto the triborheometer such that the grooves are oriented orthogonally to the sliding direction. The textured surfaces contain raised stripes with widths 25 μm ≤ $a$ ≤ 200 μm, valleys with widths 25 μm ≤ $b$ ≤ 100 μm, and height $c$ = 35 μm (Fig. 1c). Newtonian lubricants consisting of mixtures of glycerol and water provide the necessary span of viscosities (0.001 Pa·s ≤ $\eta$ ≤ 1.414 Pa·s) for obtaining the full EHL Stribeck curve. A soft poly(dimethyl siloxane) (PDMS) ball is pressed down against flat and patterned substrates at a fixed $F_N$ = 1.5 N across all experimental conditions. Sliding velocities of 500 μm/s $\leq U \leq 35$ mm/s are applied to the tribopairs, generating Reynolds numbers ranging from 4 to 98 (Supplementary Information). The geometry and flow conditions used in our study coincide with that of human fingers sliding on hard surfaces[17].

There is a critical transition in the EHL friction coefficient for patterned surfaces as $S$ increases, while this phenomenon is consistently absent in flat tribopairs (Fig. 1a). The critical friction coefficient $\mu_{c,exp}$ is defined by the local maximum in $\mu$ for a patterned geometry (Fig. 2a). Signs of these peaks have previously been observed in scratched stainless steel-PDMS tribopairs[10] and in fibrillated articular cartilage[18]. We hypothesize that the critical change in $\mu$ is due to a micro-EHL to macro-EHL transition, defined as the condition under which the lubrication film thickness jumps to maintain laminar flow at fixed normal forces (Fig. 2b). The jump in $H$ at an intermediate value of $S$ is intriguing, because it occurs for all four materials at a fixed geometrical ratio where $h_c/H_c \approx 0.1$ (Fig. 2c). The variable $h$ refers to the lubrication film thickness between the top of the texture and the contacting PDMS ball, while $H = h + c$ (Fig. 1c). In other words, the jump in the lubrication film thickness occurs at speeds where the film thickness reaches $\approx 11\%$ of the height of the stripes, independent of the type of soft material used. These observations suggest that the EHL friction is dictated solely by the physics of flow and not by interfacial interactions. In contrast, only a steady increase in $H$ is found for flat PDMS-PDMS tribopairs, in agreement with documented literature[15].

To investigate the jumps in $H$ and $\mu$ for textured tribopairs, we consider the length scales responsible for EHL flows (Fig. S2a). The total film thickness $H$ is found from the summation of three components: first, a force balance between the fluid pressure $p$ and elastic modulus $E$ generates a compression of the patterns ($p = E\varepsilon$, $\varepsilon$ is the pattern compressive strain, Fig. S2 and S3); second, effectively smooth surfaces experience an increase in film thickness as a function of sliding speed and elasticity, which is available from empirical correlations[19]; third, we add experimental differences in the film thickness relative to the zero static position, which capture the jumps for textured tribopairs at critical values of $S$ (Fig. 2b). The soft textures may undergo bending, but the estimated shear strain is comparable to the texture height and does not qualitatively change the overall physics (Fig. S4, Supplementary Information). The summed value of $H$ accounts for the elastic deformation of patterns (Fig. S6). It is used in Reynolds' theory, along with contact area measurements enabled by fluorescent dye transfer (Fig. S5, Table S1 & S2, Supplementary Information), to compute $\mu$ for micro-EHL ($S < S_c$) and macro-EHL ($S \geq S_c$).

Our framework can be understood this way: at low speeds, tribopairs experience shear and normal forces subject to the full effects of texture compression; at high speeds, the film thickness is sufficiently large such that textures are indistinguishable from an effectively flat surface. Fig. 3a shows that the separation of flow regimes using this interpretation gives rise to two lubrication scalings. In the micro-EHL regime ($S < S_c$), $F_S$ and $F_N$ are given by

$$F_N = PA \sim \frac{\eta U a A_a}{h^2} + \frac{\eta U a A_b}{H^2} \approx \frac{\eta U a A_a}{h^2} \quad \text{and} \quad F_S = \tau A \sim \frac{\eta U A_a}{h} + \frac{\eta U A_b}{H} \approx \frac{\eta U A_a}{h} \qquad (1)$$

Where $\tau$ is the shear stress, $A_a$ is the area of the raised stripes, and $A_b$ is the area of the valleys. For predicting the macro-EHL behavior at $S \geq S_c$, we use

$$F_N = PA \sim \frac{\eta U (2R) A_{total}}{h^2} \quad \text{and} \quad F_S = \tau A \sim \frac{\eta U A_{total}}{H} \qquad (2)$$

where $A_{total}$ is the total contact area. Because equations (1) and (2) are scaling relations, the prefactors $k_{micro}$ and $k_{macro}$ are used to generate exact solutions for each tribopair geometry. These prefactors scale differently as the length of the raised stripes, $a$, depending on the flow regime ($k_{micro} \sim a$ while $k_{macro} \neq f(a)$, Fig. S7). Detailed discussion of these scaling factors, along with their possible geometrical and material origins, are provided in the Supplementary Information. The predicted value of $\mu_{c,lub}$ is the average of $\mu = F_S/F_N$ at $S = S_c$, obtained from equations (1) and (2). Fig. 3b shows a representative example of the model overlaid on the experimental data for a PDMS-PDMS tribopair ($a = 100$ μm, $b = 100$ μm, $c = 35$ μm). A number of simplifying assumptions are used to enable the computation of $\mu_{c,lub}$ (Supplementary Information). Using this method, we are able to obtain the full EHL tribological phenomena for textured tribopairs with different materials and geometries (Fig. S10 to S18).

The relevance of our work to materials science and engineering is highlighted in Fig. 4: not only is it successful in obtaining $\mu_{c,lub}$ for a range of soft materials, but it also shows that $\mu_{c,exp}$ falls on a master curve as a function of texture geometry ($a$ and $b$) and reduced elastic modulus ($E'$). This latter scaling is based on purely experimental data and is free from fitting parameters. The reduced elastic modulus averages the modulus for both substrates, normalized by their Poisson's ratios (Supplementary Information). Fig. 4a shows that the predictions agree well with experimental observations for PDMS ($E = 2$ MPa, hydrophobic), mercaptoester ($E = 137$ MPa, hydrophobic), polyester ($E = 1.2$ GPa, hydrophobic) and poly(ethylene glycol) diacrylate-alginate double network hydrogel ($E = 3.4$ MPa, hydrophilic, $MW_{PEGDA} = 700$ g/mol, $MW_{alginate} = 216$ g/mol) textured substrates paired with a PDMS ball on the triborheometer. This agreement shows that the EHL tribology is fully captured for various materials despite the assumptions inherent in our model. Furthermore, Fig. 4b shows that $\mu_{c,exp}$ for the various tribopairs correlates well with the geometry ratio $a/(a+b)^{0.5}$ when the friction coefficient is normalized appropriately. The linear correlation between $\mu_{c,exp}$ and $a/(a+b)^{0.5}$ is physically explained based on dimensional scaling with respect to the contact line and the repeating unit of the geometry. Briefly, the fluid pressure on an effectively smooth surface is equivalent to the pressure on the top of the textures at the critical transition. With this correlation, we are able to bridge the value of $h$ for the smooth surface with a textured surface based on surface geometry. Combining this relation with Reynolds' equations generates the observed linear scaling between $\mu_{c,exp}$ and $a/(a+b)^{0.5}$ (Fig. S8 and S9, Supplementary Information).

Fig. 4 provides a framework for which orthogonal patterns on soft surfaces can be used to alter lubricated friction. Its broad applicability comes from the universal observation and prediction of the EHL friction for different tribopairs. Although it is known that friction is dependent on many factors such as wetting[20], surface geometry[21], applied pressure[22], and temperature[23], quantification of well-characterized tribological systems to form a uniform and experimentally accessible theory is incomplete. Simple models that identify salient structural and material properties, such as the one presented here, provide a foundation on which predictions may be further expanded to irregular textures and surfaces with random roughness. The ability to control friction is of great importance in biomedical applications such as joint implants, where low friction is desired with specific pressures and velocities[7,24]. Friction is also important in the bulk mechanics of particulate suspensions[25-27] and in the design of food and cosmetic products[28,29] as they feature deformable structures that slide against one another.

**Acknowledgements:** The authors thank John F. Brady, Ronald G. Larson, Joelle Frechette, and Alison Dunn for scientific discussions.

**Author contributions:** Y. P. and L. C. H. designed the study, developed and validated the theory, and wrote the paper. Y. P., C. M. S., and C. N. H. designed and conducted experiments. All authors were funded by North Carolina State University startup funds and the AAAS Marion Milligan Mason Award.

**Competing Interests:** The authors declare no competing financial interests.

**Methods**

Detailed descriptions of the materials and methods used are provided in the Supplementary Information.

Fabrication of microtextured polymer surfaces

Stencils (silicon wafers, $R$ = 76.2mm) for pattern transfer are produced by standard UV lithography. Polydimethylsiloxane (PDMS; Sylgard 184, Dow Corning) is prepared at a 10:1 base/curing agent mass ratio (total of 12 g) and cast onto patterned Si wafers. The samples are then heated at 70°C overnight for curing. The cured PDMS samples are cut into 0.6 cm × 1.5 cm rectangular slabs for tribological measurements.

Microtextured surfaces made from mercaptoesters, polyesters, and double network (DN) hydrogel are produced via replica molding, where the patterned PDMS is the stencil. Microtextured PDMS stencils are produced by the above-mentioned procedure. Crosslinked mercaptoesters are produced from Norland Optical Adhesive (NOA) 65 (Norland Products). We pour 12 g of NOA 65 into the patterned PDMS mold and allow it to rest until the bubbles dissipate. The sample is exposed to UV light ($\lambda$ = 254 nm) for two hours to ensure complete crosslinking. Crosslinked polyesters are produced by mixing the base (Clear-Lite Casting Resin) with methyl ethyl ketone peroxide catalyst (MEKP) (TAP Plastics) at 40:1 w/w %. A total of 12 g of polyester resin is poured over the patterned PDMS mold and left at room temperature overnight before being heated at 70°C for one hour. Patterned mercaptoester and polyester are cut into 0.6 cm × 1.5 cm rectangular slabs for tribological measurements.

Poly(ethylene glycol) diacrylate (PEGDA)/alginate double network (DN) hydrogels are prepared in 2 steps. First, PEGDA (Sigma-Aldrich) and sodium alginate (Sigma-Aldrich) are mixed with deionized water at 5:95 w/w% and 40:60 w/w% respectively. The PEGDA and alginate solutions are then mixed at 10:1 w/w%. The photoinitiator Darocur 1173 (Sigma-Aldrich) is added to the mixture at 0.5:99.5 w/w %. We remove the bubbles in the mixture by tumbling at 25 rpm for 24 hours and then centrifuging at 10000 rpm for 20 minutes. Following the bubble removal step, 12 g of the DN hydrogel precursor liquid is poured over the patterned PDMS and cured under UV light for one hour ($\lambda$ = 254 nm). Once completely crosslinked, the hydrogel sample is removed from the PDMS mold and soaked in a 1 M calcium chloride solution for 24 hours. DN hydrogel samples for tribological testing are cut into 0.6 cm × 1.5 cm rectangular slabs.

Tribological measurements

The tribology experiments are conducted with a ball-on-three-plates geometry on a stress-controlled rheometer (DHR-2, TA Instruments) at a temperature of 20°C. Tests are conducted with normal force $F_N$ = 1.5 N and in the presence of lubricants consisting of water-glycerol mixtures. The sliding velocity between the ball and the plates ranges from 1 rad/s to 80 rad/s.

Contact area measurements

Static contact areas are measured using a Leica TCS SP8 inverted microscope equipped with a 10 × dry objective. A fluorescently dyed PDMS sphere is pressed onto the microtextured polymer surfaces at $F_N$ = 1.5 N. The resultant fluorescent contact area on the textured surfaces is quantified by image processing (Fig. S5).

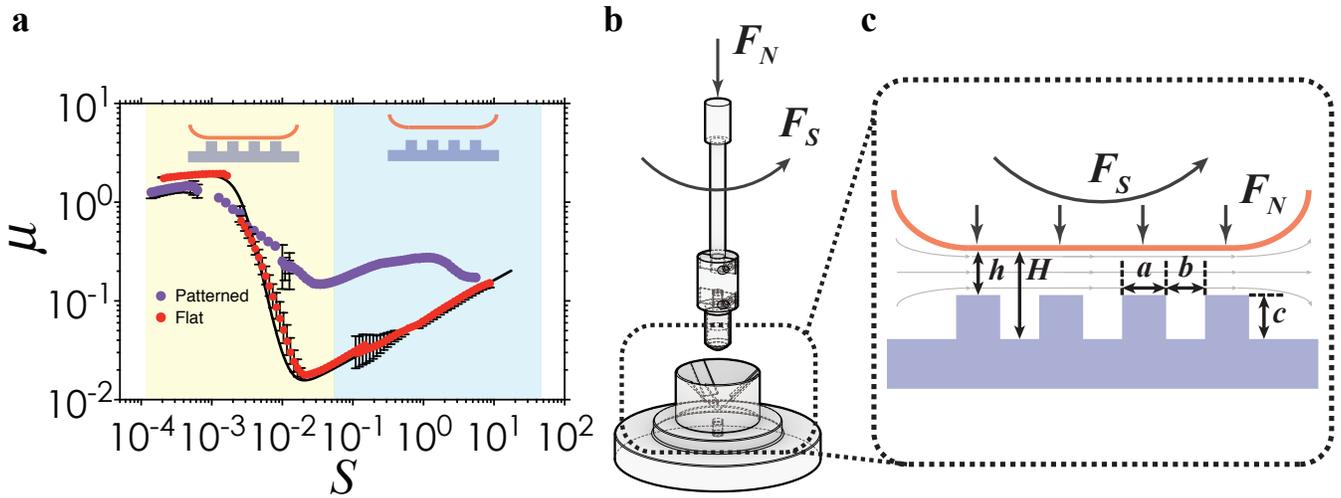

**Figure 1 | The Stribeck curves for flat and patterned soft materials. a**, Steady state sliding friction as a function of the Sommerfeld number $S$ for flat (purple circles) and patterned (red circles) PDMS-PDMS tribopairs with a Newtonian lubricant. Solid black line represents an empirical fitting for the Stribeck curve of flat PDMS-PDMS tribopairs. The pale yellow region indicates boundary and mixed lubrication, where two surfaces are relatively close to each other with direct touching. The pale blue region indicates the EHL regime. Error bars indicate the standard deviation from 6 independent measurements. Each Stribeck curve is obtained by independent measurement of 3 sets of substrates with different lubricant viscosities. The inset in the pale yellow area indicates a direct contact between two opposing surfaces while the inset in the pale blue region shows a full separation between surfaces. **b**, Ball-on-three-plate triborheological accessory, in which the top ball rotates and slides against three bottom substrates at a constant normal force $F_N$. The friction force $F_S$ is obtained by converting the torque detected at increasing sliding speeds. **c**, A schematic of the tribopair contact, in which the orange solid line represents the top PDMS ball. Solid purple represents the bottom textured substrate with streamlines indicating the direction of lubricant flow.

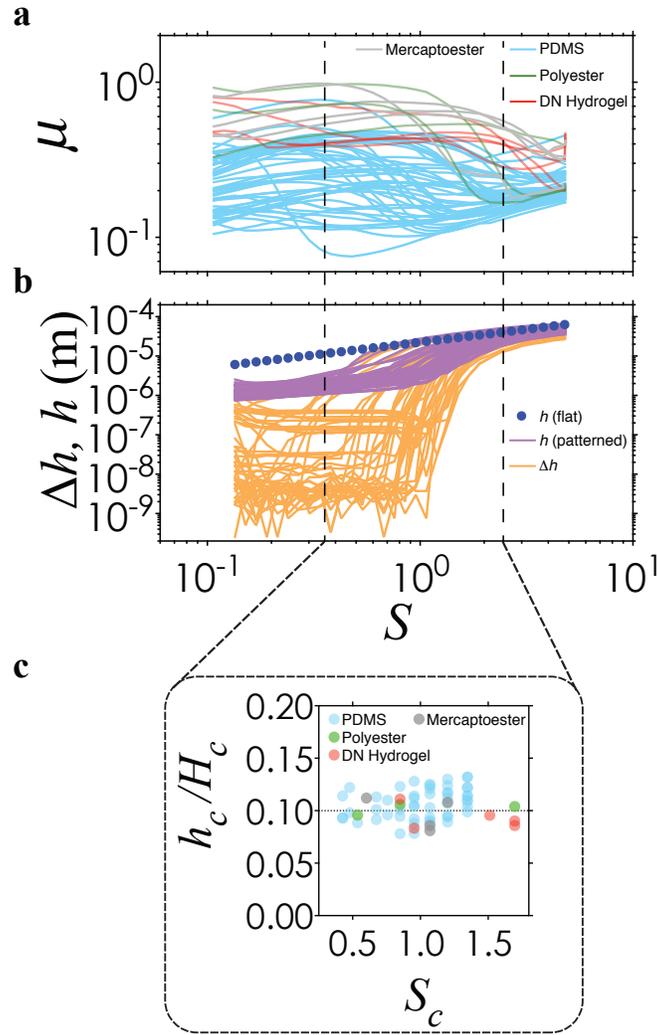

**Figure 2 | EHL friction and lubrication film thickness. a**, Steady state tribology of patterned substrates in the EHL regime: PDMS (blue), polyester (green), mercaptoester (gray), and double network hydrogel (red), where each curve of the same color represents the same material but with different geometries. Error bars are too small to show. **b**, Fluid film thickness plotted as a function of $S$ for patterned PDMS substrates, where each curve of the same color represents different geometries. The experimentally measured $\Delta h$ values (orange solid lines) show a significant increase at a critical point and it contributes to a significant increase of the total fluid film thickness $h$ at critical values of $S$ (pink solid lines). The value of $h$ increases linearly with $S$ for flat PDMS substrates (blue filled circles). Vertical dashed lines across a and b indicate the range of $S$ in which critical transitions are observed for all substrates tested in this study. **c**, The ratio of $h_c$ and $H_c$ plotted against the critical $S_c$ for PDMS (light blue circles), polyester (green circles), mercaptoester (gray circles) and double network hydrogel (red circles). Horizontal dotted line indicates that $h_c/H_c \approx 0.10$ is observed for all materials and geometries.

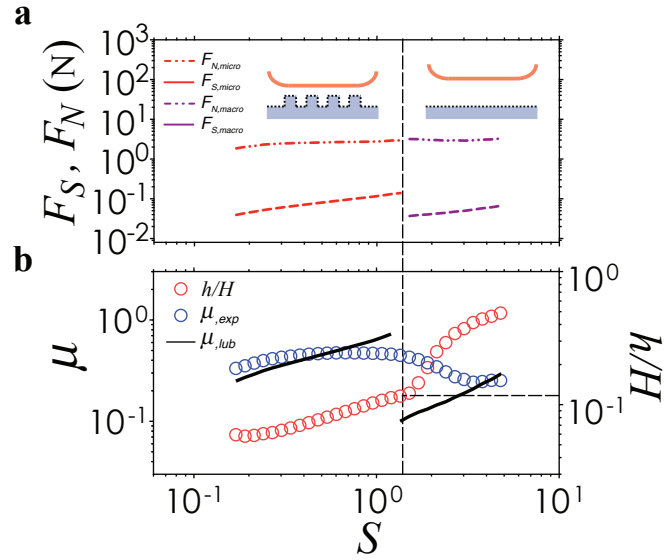

**Figure 3. | Predicting the critical EHL transitions for patterned geometries. a**, The predicted friction force (dashed lines) and predicted normal force (dash-dotted lines), separated by the vertical dashed line into the micro-EHL (red) and macro-EHL regimes (purple). Inset: Schematic of the physics of textural compression at small and large $S$. **b**, The experimental data for $\mu$ (blue open circles) and $h/H$ (red open circles) is plotted as a function of $S$. Horizontal dashed line indicates the critical value of $h_c/H_c = 0.103$. Black solid lines are for the predicted EHL friction coefficient. The data in **a** and **b** are shown for a representative PDMS patterned tribopair where $a = 100$ μm, $b = 100$ μm, and $c = 35$ μm.

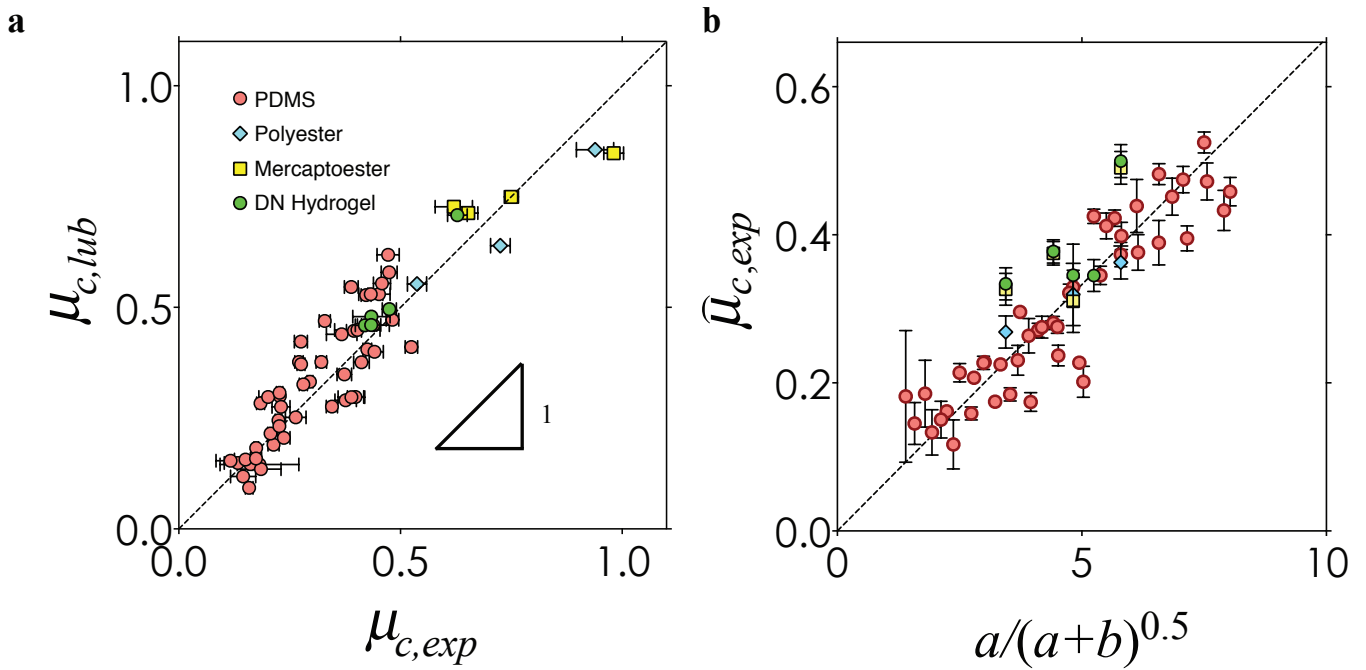

**Figure 4 | Material- and geometry-based model for the critical EHL friction coefficient. a**, The predicted critical $\mu$ agrees well with the experimental data for patterned substrates. **b**, The normalized critical $\mu$ shows a linear relation with the geometrical parameter $a/(a+b)^{0.5}$ for all patterned surfaces with different materials. In **a** and **b**, the substrates used are PDMS (red circle), mercaptoester (yellow square), polyester (blue diamond), and double network hydrogel (green circle). Error bars represent standard deviations from 6 independent samples.

# Supplementary Information

## Lubricated friction of textured soft substrates

Yunhu Peng, Christopher M. Serfass, Catherine N. Hill, Lilian C. Hsiao[*]

Department of Chemical and Biomolecular Engineering, North Carolina State University, Raleigh, NC 27695, USA

Corresponding author.

*Email: lilian_hsiao@ncsu.edu

Table of Contents

# Materials and Methods

## Fabrication of microtextured PDMS

Micropatterned silicon wafers (wafer radius $R$ = 76.2 mm) used to generate the model substrates are produced by standard lithography. Photomasks of patterns are designed in AutoCAD and printed by FineLine Imaging. The striped patterns produced by the photomasks onto the etched wafers have dimensions in which both $a$ and $b$ ranges from 25 μm to 200 μm. Silicon wafers are first cleaned by a plasma cleaner (Diener Electronic) for 1 minute. Then, OmniCoat (MicroChem) is added dropwise onto the wafer in a spin coater operating at 3000 rpm. The coated wafer is then placed on a hot plate at 200 °C for 1 minute. The OmniCoat serves to enhance the bonding between the silicon wafer and SU-8 photoresist. After the wafers are completely cooled to room temperature, a layer of roughly 10 g SU-8 2050 (MicroChem) is deposited on top of it by spin coating for 1 minute at 3000 rpm. The wafer is removed and heated at 65°C for 1 minute, followed by another heating step at 95°C for 7 minutes. The purpose of prebaking is to remove excess solvent from the SU-8 2050. The prebaked wafer is placed inside a Suss MA6/BA6 Contact Aligner and exposed under ultraviolet light ($\lambda$ = 400 nm) for 10 seconds with the printed photomask on top of the wafer. The wafers are removed from the aligner and cured at 65°C for 1 minute followed by 95°C for 6 minutes. The cooled wafers are developed with 1-methoxy-2-propanol acetate and then cleaned with pure isopropanol. The resulting textured surfaces serve as molds for the fabrication of PDMS textured substrates.

PDMS textured surfaces are prepared by pouring 12 g Sylgard 184 (Dow Corning) at a base to curing agent ratio of 10:1 w/w% onto the wafers and curing at a temperature of 70°C overnight. The cured substrate has a Young's modulus of 2 MPa and a thickness of 1.9 mm[1].

They are cut into 0.6 cm × 1.5 cm rectangular slabs for use in tribological characterization. The same curing process is used to fabricate spherical PDMS balls with a radius of 1.27 cm in a custom stainless steel mold.

Fabrication of microtextured mercaptoester and polyester

Microtextured surfaces made from mercaptoesters and polyesters are synthesized for tribological testing. Both materials have higher elasticity than PDMS. These surfaces are produced via replica molding, a process for shaping materials in which an intermediate mold is used to transfer a pattern onto the desired material[2-4]. Two materials are employed: polyester resin and mercaptoester. The elastic modulus for polyester and mercaptoester are 1.2 GPa[5] and 137 MPa[6] respectively. Before casting, a micro-textured PDMS mold is created as described previously. Polyester resin (Clear-Lite Casting Resin) contains styrene monomer and we use methyl ethyl ketone peroxide (MEKP) (TAP Plastics) as the catalyst. Mercaptoester is obtained from Norland Optical Adhesive (NOA) 65 (Norland Products). We use a Mineralight XX-20S UV Bench Lamp (UVP) ($\lambda$ = 254 nm) to cure the material. The resin is mixed with styrene monomer in a 40:1 mass ratio. Following this step, one drop of MEKP catalyst is added for every 4 g of polyester resin used. The mixture is then mixed and degassed. For polyester substrates, 12 g of resin is poured over the PDMS mold and left at room temperature overnight before being treated at 70°C for one hour prior to removal from the mold. Slabs for tribological testing are then cut to size using a vertical bandsaw. For mercaptoester substrates, 12 g of NOA 65 is poured over the PDMS mold, covered in aluminum foil, and allowed to sit until all bubbles dissipate. The NOA 65 is then exposed to UV light for one hour before being flipped and exposed for another hour. The cured mercaptoester is removed from the mold. Slabs for tribological testing are cut using a razor blade.

Fabrication of microtextured double network (DN) hydrogel

Microtextured poly(ethylene glycol) diacrylate (PEGDA) and alginate DN hydrogel substrates are similarly made from replica molding. In order to prepare the patterned hydrogel, the PEGDA and alginate monomers are crosslinked in two steps. Alginic acid and PEGDA are obtained from Sigma-Aldrich and used without further purification. Both alginate and PEGDA are mixed with deionized water at a concentration of 5% and 40% respectively. A mixture of the monomers is obtained by mixing PEGDA and alginate monomer solutions at 10:1 mass ratio. The photoinitiator 2-hydroxy-2-methylpropiophenone or Darocur (Sigma-Aldrich) is added to the mixture at 0.5:99.5 w/w %. The mixture is stirred for 10 minutes before being transferred to a glass vial. The entire container is covered in aluminum foil and allowed to tumble at 25 rpm for at least 24 hours. After 24 hours, the sample is placed in a centrifuge at 10000 rpm for 20 minutes. The sample is then poured slowly, to prevent entrapment of bubbles, into the middle of the PDMS mold and placed under UV light for one hour ($\lambda$ = 254 nm). Once completely crosslinked, the hydrogel sample is removed from the PDMS mold and soaked in a 1 M calcium chloride solution for 24 hours. DN hydrogel samples for tribological testing are cut using a razor blade.

Tribological characterization

The tribology experiments are conducted with a ball-on-three-plates geometry on a stress-controlled rheometer (DHR-2, TA Instruments) at a temperature of 20°C. The geometry has a ball attached at the top and three plates inserted in the bottom tray (Figure 1b). The ball is lowered to make contact with the three plates at a fixed normal force of $F_N$ = 1.5 N in the presence of a lubricant. The sliding direction of the ball is perpendicular to the stripes. The lubricant consists of a mixture of deionized water and glycerol (Sigma-Aldrich, ≥ 99.5%). The water-glycerol mixtures are prepared at a ratio of 0:100 w/w %, 30:70 w/w%, 10:90 w/w%, and 100:0 w/w% ($\eta$ = 0.001 Pa·s, 0.024 Pa·s, 0.768 Pa·s, and 1.412 Pa·s, respectively). The relative sliding velocity ω between the ball and the plates ranges from 1 rad/s to 80 rad/s. We collect 40 data points at 40 different sliding velocities within this range. Each data point is obtained by averaging the values obtained over 25 seconds.

# Lubrication analysis in thin film gap between two smooth surfaces

Lubrication refers to the flow of fluids through thin gaps separated by two solid surfaces. A schematic of lubricated flow is shown in Figure S1, and here we describe the well-known Reynold's equations used to predict the forces involved in such flows for the sake of clarity[7].

When the top surface is sheared against a smooth bottom surface, the average relative velocity between two surfaces is $U$. In our experiments, the normal force $F_N$ is constant. Lubricant flows between the two surfaces at a volumetric flow rate $q$. The gap height between the top and bottom smooth surfaces is $h_{smooth}$, which is a function of $x$ and therefore can be expressed as $h_{smooth} = h_{smooth}(x)$[8]. The entrance of the lubrication regime starts at $x = x_1$ and ends at $x = x_2$. The pressure at the entrance and the exit are $p_1$ and $p_2$ respectively. The lubricant velocity is given by $u = u(x, y)$. Using the equations of motion, the local velocity of the lubricant flow can be expressed as the following:

$$u_x(x,y) = U[1-(\frac{y}{h_{smooth}})] - \frac{h^2}{2\eta}\frac{dp}{dx}[(\frac{y}{h_{smooth}}) - (\frac{y}{h_{smooth}})^2] \quad (S1)$$

$$u_y(x,y) = 2\frac{dh}{dx}[(\frac{3q}{h_{smooth}}) - U][(\frac{y}{h_{smooth}})^2 - (\frac{y}{h_{smooth}})^3] \quad (S2)$$

The flow rate $q$ in this 2D model can be obtained by integrating $u_x$ over the gap height:

$$q = \int_0^{h_{smooth}} u_x dy = \frac{Uh_{smooth}}{2} - \frac{h_{smooth}^3}{12\eta}\frac{dp}{dx} \quad (S3)$$

where $\eta$ represents the fluid viscosity. By rearranging the above equation, we can solve for the pressure gradient:

$$\frac{dp}{dx} = \frac{6U\eta}{h_{smooth}^2} - \frac{12q\eta}{h_{smooth}^3} \quad (S4)$$

Integrating (S4) provides the pressure in the thin gap as:

$$p(x) - p(x_1) = 6U\eta \int_{x_1}^{x} h_{smooth}^{-2} dx - 12q\eta \int_{x_1}^{x} h_{smooth}^{-3} dx \quad (S5)$$

When we substitute $x_2$ into $x$, we obtain the pressure drop $\Delta p$,

$$\Delta p = 6U\eta \int_{x_1}^{x_2} h_{smooth}^{-2} dx - 12q\eta \int_{x_1}^{x_2} h_{smooth}^{-3} dx \quad (S6)$$

The resulting force components on this plane are thus:

$$F_S = -\int_{x_1}^{x_2} \left(\frac{dh_{smooth}}{dx} p + \eta \frac{\partial u_x}{\partial y}\bigg|_{y=h}\right) dx \quad (S7)$$

$$F_N = \int_{x_1}^{x_2} p \, dx \quad (S8)$$

Equation (S5) indicates that the pressure $p \sim \frac{U\eta L}{h^2}$, where $L = x_2 - x_1$ and $h$ is the representative gap height. The value of $h$ can be estimated as an average of $h_{smooth}$ and is expressed with the following term:

$$h = \frac{\int_{x_1}^{x_2} h_{smooth} dx}{x_2 - x_1} \quad (S9)$$

The shear stress in the lubrication system is:

$$\tau_{xy} = \eta \frac{\partial u_x}{\partial y} \quad (S10)$$

Equation (S10) indicates that the scaling relation for shear stress is $\tau_{xy} \sim \frac{U\eta}{h}$. The scaling equation for pressure and shear stress indicates that the ratio of two stresses follows the scaling:

$$\frac{\tau_{xy}}{p} \sim \frac{h}{L} \quad \text{(S11)}$$

The value of $h$ is typically in the range of microns for most engineering applications, while $L$ in our case is around the millimeter range. The ratio of those two parameters is

$$\frac{h}{L} \sim O(10^{-3}) \quad \text{(S12)}$$

This small ratio of $h$ to $L$ justifies the use of lubrication approximation in our study. Using the scaling relations for both stresses, we estimate the forces in the experimental system for flat tribopairs as

$$F_N \sim \frac{U\eta L}{h^2} A \quad \text{(S13)}$$

$$F_S \sim \frac{U\eta}{h} A \quad \text{(S14)}$$

where $A$ represents the total contact area.

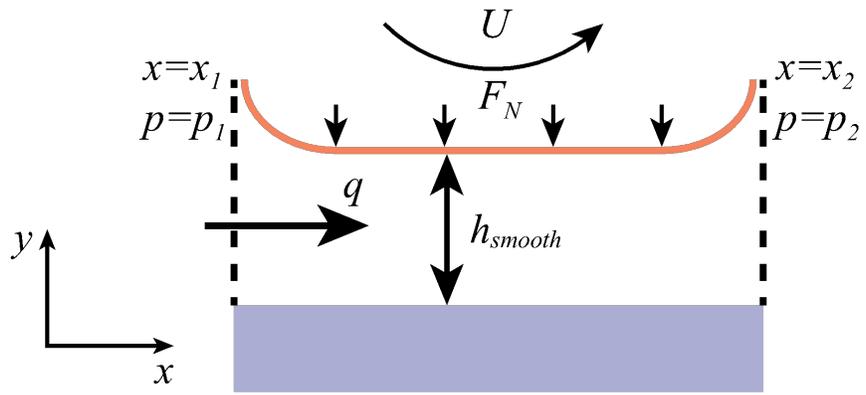

Figure S1| A model system to illustrate thin film lubrication between two smooth surfaces.

# Deformation of striped PDMS surfaces

Two types of deformations exist in the system: a vertical strain and a horizontal strain as shown. In our predicted physical model, we consider both types of deformation and their relative influence on tribological behavior.

## Compression of the stripes

The thin fluid film between two surfaces generates a normal pressure which separates the two surfaces from each other and also results in the compression of the bottom stripes. Figure S2 shows the side view of this type of localized compressive deformation. We neglect the deformation of the bulk PDMS of the bottom plate because the deformation of the bulk (Figure S3) is minimal compared to the compression of the textures. The fluid pressure exerted on the textures changes the overall fluid film thickness and subsequently alters the tribological behavior of textured surfaces. We compute the deformation from

$$\frac{U\eta a}{\left(h_a + c \times \varepsilon\right)^2} = E\varepsilon \quad (S15)$$

where $h_a$ is gap height altered from empirical correlations available in the tribology literature for smooth, flat tribopairs[9]. In this equation, $E$ is the elastic modulus of the bulk material. For PDMS, $E \approx 2$ MPa. For PEGDA/Alginate hydrogel, $E \approx 3.4$ MPa. For NOA, $E \approx 137$ MPa. For polyester, $E \approx 1.2$ GPa. The compressive strain $\varepsilon$ on the stripe is estimated to be less than 2% throughout all of our experiments.

Shear strain of the stripes

The shear force applied on the stripes causes a bending strain in the direction of the flow as shown in Figure S4. The shear stress $\tau$ applied follows the equation $\tau = G\varepsilon'$ where $G$ is the shear modulus and $\varepsilon'$ is the shear strain. The shear modulus $G$ can be calculated based on the elastic modulus $E$ and Poisson ratio $v$ where $G = \dfrac{E}{2(1+v)}$. We estimate that the maximum $\varepsilon'$ is 10% and the maximum angle is $\alpha = 16.6°$. With the max bending angle and the max $\alpha$, we can calculate the actual texture height of the texture $c_{act}$ from Pythagorean theorem where $c_{act} = \sqrt{c^2 - (a \times \varepsilon')^2} = 33.54 \mu m$.

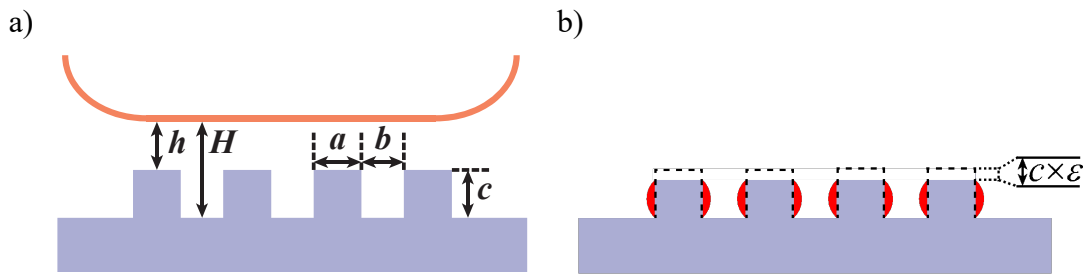

Figure S2| (a) Side view of the experimental setup and the surface geometrical parameters. (b) Deformation of the bottom texture surface due to the lubricant pressure.

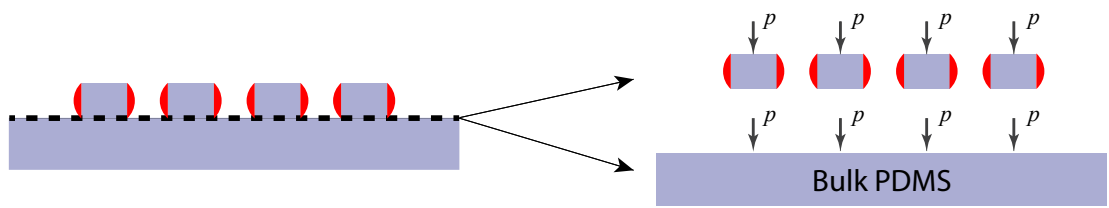

Figure S3| Deformation on the striped PDMS can be considered as two separate parts: the bulk PDMS and stripes. Dash line represents the cut off interface of two separate parts. Overall pressure exerted on the stripes and bulk PDMS are the same.

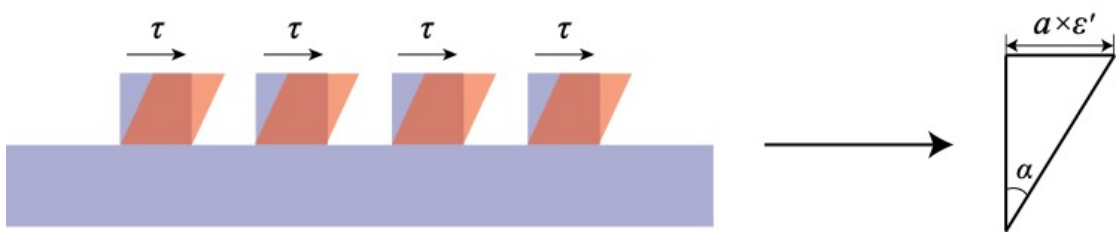

Figure S4| Shear bending of the striped surface under lubricated sliding. $\tau$ is the shear stress. $\alpha$ is the bending angle.

# The reduced Young's modulus of tribopairs

The reduced Young's modulus $E'$ is defined as

$$\frac{1}{E'} = \frac{1}{2}\left[\frac{1-v_1^2}{E_1} + \frac{1-v_2^2}{E_2}\right] \quad \text{(S16)}$$

where $E_1$ and $E_2$ are the elastic moduli for the two individual surfaces in a tribopair. The values $v_1$ and $v_2$ are the Poisson's ratios for each surface. For a PDMS-PDMS tribopair, $E_1 = E_2$ & $v_1 = v_2$. Therefore, the reduced modulus of a PDMS-PDMS tribopair is

$$E'_{PDMS-PDMS} = \frac{E_{PDMS}}{1-v_{PDMS}^2}. \quad \text{(S17)}$$

For PDMS-mercaptoester & PDMS-Polyester tribopairs, the elastic modulus of PDMS is much smaller than both mercaptoester and polyester and therefore $\frac{1-v_2^2}{E_2} \ll \frac{1-v_{PDMS}^2}{E_{PDMS}}$ (material properties for mercaptoester and polyester are represented by $E_2$ and $v_2$). With this relation, $E'$ for PDMS-mercaptoester & PDMS-Polyester tribopairs both have the form

$$E' \approx \frac{2E_{PDMS}}{1-v_{PDMS}^2} \quad \text{(S18)}$$

Equation S18 shows that the reduced modulus of PDMS-mercaptoester & PDMS-Polyester tribopairs are similar. Using equations S16, S17 & S18, we compute the reduced Young's moduli for the PDMS-PDMS, PDMS-mercaptoester, PDMS-Polyester & PDMS-PEGDA/Alginate DN Hydrogel tribopairs as 2.67 MPa, 5.33MPa, 5.33MPa and 3.36MPa respectively.

# Reynolds Number (Re) during sliding conditions

Our model assumes that the fluid flow in our lubricated system is laminar. To validate this assumption, we compute the Reynolds Number (Re) for our system over the range of sliding velocities used. The Re number is $\text{Re} = \frac{\rho U L}{\eta}$, where $\rho$ is the density of the lubricant, $U$ is the average sliding velocity, $L = 2R$ is the characteristic length of the contact, and $\eta$ is the lubricant viscosity. In our experiments, we use mixtures of water and glycerol as the lubricants. The viscosities of the mixture are reported in the Materials and Methods section. If we define the radius of the PDMS ball ($R = 0.0127$m) as the characteristic length, then $\text{Re} \in [4, 98]$. This suggests that laminar flow is maintained throughout all experiments.

# Contact area of striped surfaces

The contact areas on textured surfaces are obtained by compression testing using the triborheometer as discussed in the Materials and Methods section. The compression tests are performed without any lubricant in static conditions. An illustration of the setup is shown in Figure S5a. Fluorescent Nile Red (Sigma-Aldrich) dissolved in toluene (0.5:99.5% w/w%) is used to dye PDMS. A dyed PDMS sphere is pressed on top of the microtextured substrates at a constant normal force $F_N$ = 1.5 N to allow dye transfer. The resultant fluorescent contact areas are quantified using confocal microscopy and image processing. Confocal microscopy images are acquired using a Leica TCS SP8 inverted microscope with a 10× dry objective with tile stitching. Confocal imaging shows that the contact area of the striped surface is within a circle indicated by the dashed line in Figure S5b. We acknowledge that the static contact area obtained with our experimental setup is different from the actual contact area under lubricating sliding conditions[10]. Although the shape of contact area is not completely circular during sliding, we make a first order assumption that the effective contact area during sliding lubrication should be close to the area measured under dry static contact because of the same normal force and overall fluid pressure. The confocal image indicates that the applied pressure is concentrated on the stripes during contact, which qualitatively agrees with the analysis of pressure distribution in the micro-EHL regime. The contact areas within the dashed circle for different striped surfaces are summarized in Table S1 & S2. We define this area enclosed in the dashed line as $A$ and the area of direct contact as $A_a$ (red area in Figure S5b). The remaining dark area in the circle is $A_b$. Considering $A$, $A_a$ and $A_b$ together, we have $A = A_a + A_b$. Using the geometrical parameters defined in Figure S2a, we obtain a relation between the contact area and total area: $\frac{A_a}{A} \approx \frac{a}{a+b}$ and $\frac{A_b}{A} \approx \frac{b}{a+b}$.

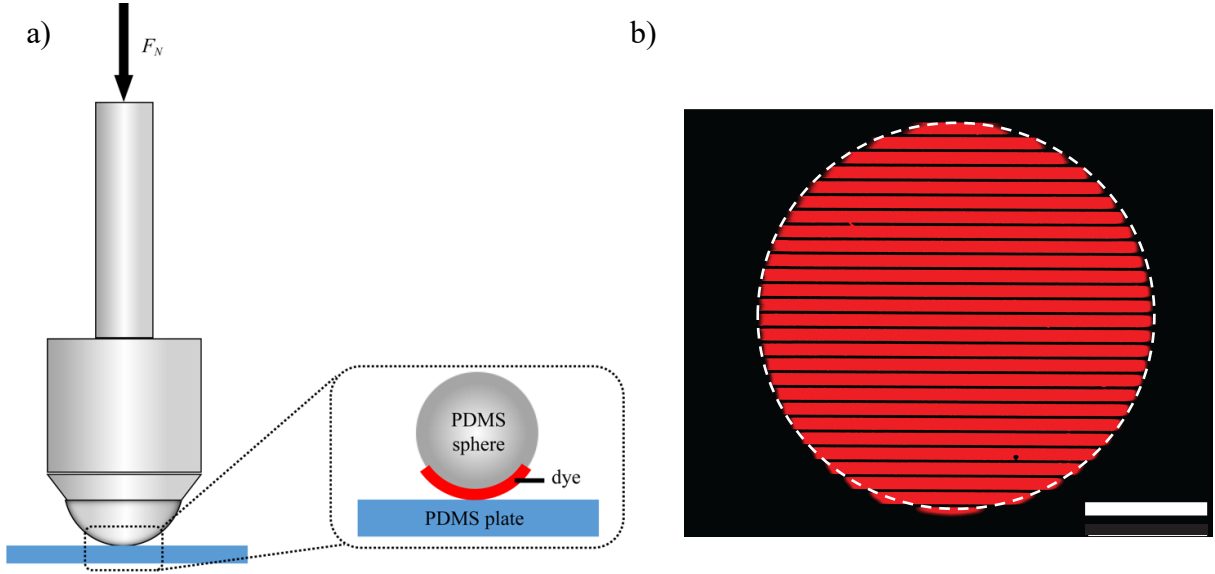

Figure S5 | Demonstration of compression test. (a) Side view of the setup for compression tests. (b) Confocal image of compressed striped surface with $F_N = 1.5$N. Scale bar represents 1mm.

Table S1 | Contact area for PDMS striped surface

| a (μm) | b (μm) | Area (m²) | a (μm) | b (μm) | Area (m²) | a (μm) | b (μm) | Area (m²) |
|---|---|---|---|---|---|---|---|---|
| 25 | 15 | 0.000009539 | 45 | 15 | 0.000008708 | 75 | 15 | 0.000009789 |
| 25 | 25 | 0.000010139 | 45 | 25 | 0.000009684 | 75 | 25 | 0.000009789 |
| 25 | 35 | 0.0000102 | 45 | 35 | 0.000009361 | 75 | 35 | 0.0000102 |
| 25 | 45 | 0.000010309 | 45 | 45 | 0.000010047 | 75 | 45 | 0.000010447 |
| 25 | 55 | 0.00001035 | 45 | 55 | 0.000010595 | 75 | 55 | 0.000009849 |
| 25 | 75 | 0.000010679 | 45 | 75 | 0.00001055 | 75 | 75 | 0.000010177 |
| 25 | 100 | 0.000010861 | 45 | 100 | 0.000010401 | 75 | 100 | 0.000010203 |
| 35 | 15 | 0.000009539 | 55 | 15 | 0.000010552 | 100 | 15 | 0.000009824 |
| 35 | 25 | 0.000010003 | 55 | 25 | 0.00001008 | 100 | 25 | 0.000008994 |
| 35 | 35 | 0.000010668 | 55 | 35 | 0.00000995 | 100 | 35 | 0.000009657 |
| 35 | 45 | 0.0000101 | 55 | 45 | 0.000010075 | 100 | 45 | 0.000010134 |
| 35 | 55 | 0.000010248 | 55 | 55 | 0.000009674 | 100 | 55 | 0.000009595 |
| 35 | 75 | 0.000010517 | 55 | 75 | 0.000010764 | 100 | 75 | 0.000009476 |
| 35 | 100 | 0.000011451 | 55 | 100 | 0.000011615 | 100 | 100 | 0.000009491 |

Table S2 | Contact area for PEGDA/Alginate DN Hydrogel, mercaptoester and polyester striped surface

| PEGDA/Alginate DN Hydrogel | | | Mercaptoester | | | Polyester | | |
|---|---|---|---|---|---|---|---|---|
| $a$ (μm) | $b$ (μm) | Area (m$^2$) | $a$ (μm) | $b$ (μm) | Area (m$^2$) | $a$ (μm) | $b$ (μm) | Area (m$^2$) |
| 55 | 35 | $9.09813 \times 10^{-06}$ | 55 | 35 | $7.20719 \times 10^{-06}$ | 55 | 35 | $5.74425 \times 10^{-06}$ |
| 55 | 55 | $1.00665 \times 10^{-05}$ | 55 | 75 | $7.48977 \times 10^{-06}$ | 55 | 75 | $5.78977 \times 10^{-06}$ |
| 55 | 75 | $9.59845 \times 10^{-06}$ | 55 | 100 | $7.74024 \times 10^{-06}$ | 55 | 200 | $6.55808 \times 10^{-06}$ |
| 55 | 100 | $1.02882 \times 10^{-05}$ | 55 | 200 | $8.27414 \times 10^{-06}$ | | | |
| 55 | 200 | $1.21445 \times 10^{-05}$ | | | | | | |

# Detailed assumptions of the physical model

1. We assume there is no solid-solid contact between the textured surfaces and the PDMS ball because EHL is the regime where two surfaces are fully separated by hydrodynamic forces.

2. Because $a\,\&\,b \in [25\mu m, 200\mu m]$ is in the order of microns, the molecular roughness in the nanometer range can be neglected.

3. The compression in our experiment is elastic because the deformation in our experiments at 2% is much lower than the yield strain of PDMS at around 60%[1]. (See the deformation of PDMS section.)

4. We assume that the top surface is mostly parallel to the textured surface ($\frac{\partial h}{\partial x}\big|_{y=h} \approx 0$). This is a valid assumption because: (1) the top PDMS sphere experiences fluid pressure which results in the flattening of its curvature; and (2) the length scales of the textures ($a \in [25\mu m, 200\mu m]$) and the contact lines (~1.9 mm) are significantly smaller than the diameter of the PDMS sphere (1.27 cm).

5. We assume that the fluid is incompressible with a constant viscosity $\eta$ because the lubricant we used in this series of experiments is a mixture of glycerol and water. Both molecules are extremely small (water 2.75 Å, glycerol 4.35 Å) compared to the size of the gap in the EHL regime (~ μm).

6. We assume there is no cavitation in the system because of the low Re and the lack of bubbles after testing.

7. The flow in the thin gap is unidirectional ($\frac{\partial u}{\partial x} = \frac{\partial u}{\partial z} = 0$, where $u = u(x,y,z)$).

8. No slip boundary conditions apply at both the peak and the valley of stripes ($\frac{\partial u}{\partial y}\big|_{y_{peak}} = \frac{\partial u}{\partial y}\big|_{y_{valley}} = 0$).

## Scaling analysis of forces in lubricated system

The friction coefficient $\mu$ is determined by the ratio of shear and normal forces and therefore we derive lubrication equations for the forces in both micro-EHL and macro-EHL regime based on our physical model. The scaling equation S13 and S14 are combined with the measured contact areas in the previous section to provide the following relations:

$$F_N \sim \frac{U\eta a}{h^2} A_a + \frac{U\eta b}{H^2} A_b \quad (S19)$$

$$F_S \sim \frac{U\eta}{h} A_a + \frac{U\eta}{H} A_b \quad (S20)$$

Equations S19 & S20 indicate that $h$ and $H$ contribute synergistically to the forces. The scaling analysis reveals that the normal force in the lubricated system is inversely proportional to $h^2$ while the shear (friction) force is inversely proportional to $h$. However, this gap height $h$ is extremely small in the lubricated system, ranging from a few hundred nanometers to a few microns, which suggests that that even a small change of $h$ will result in a huge difference in both $F_N$ and $F_S$. Therefore, the fluid film thicknesses ($h$ and $H$) are essential in determining forces in lubrication.

## Obtaining the total lubrication film thickness

To obtain the gap height $h$, we sum three components in our system: (1) the change in $h$ as a function of $S$ and other material properties for flat tribopairs, to represent the bulk substrate before the consideration of textures; (2) the compressive strain on the stripes caused by the increasing fluid pressure as a function of $S$; and (3) the experimentally measured change in $h$ for microtextured surfaces across all *S*. The first component involves using an existing empirical equation appropriate for our experimental system. The empirical equation we use has been adopted to various soft tribological systems previously[9,11,12]:

$$h_{smooth} = 11.15(U\eta)^{\frac{2}{3}} F_N^{-\frac{2}{9}} E'^{-\frac{4}{9}} \left[\frac{R_{ball}}{2}\right]^{\frac{7}{9}} \left[1 - e^{0.72k}\right] \quad (S21)$$

where *E'* is reduced Young's modulus of the tribopair, $R_{ball}$ is the radius of the PDMS sphere and *k* is the ratio of semi-major axis to semi-minor axis of the elliptical contact area. In our analysis, *k* = 1 for all contacts. Equation (S21) is slightly modified to account for the pressure on the stripes. At the onset of EHL for the smooth surfaces, $h_{smooth}$ is within a few micrometer range (~ μm). It is reasonable to expect that *h* is within the same range for textured surfaces. Comparing the magnitude of *h* with $c = 35 \mu m \approx H - h$, it is seen that $c \gg h$ and thus $H \gg h$. With this relation, the two terms in S19 obey the relation $\frac{U\eta a}{h^2} A_a \gg \frac{U\eta b}{H^2} A_b$ which supports our assumption that the fluid pressure is mostly concentrated on the raised stripes. Since we maintain $F_N$ = 1.5N, we have:

$$p_{smooth} A_{smooth} \approx p_a A_a . \quad (S22)$$

Rearranging equation S22, we get:

$$h_a = h_{smooth} \sqrt{\frac{A_a}{A_{smooth}} \times \frac{a}{2R}} \quad (S23)$$

where $h_a$ is the predicted gap height between the top surface and the stripe, $h_{smooth}$ is the fluid film thickness with smooth surfaces predicted by equation S21, $A_{smooth}$ is the contact area with smooth surfaces and 2*R* is the diameter of the contact area. Both $A_{smooth}$ and 2*R* are obtained directly from the compression experiments on a flat surface. Equation S23 is the modified empirical equation for our predicted fluid film thickness. This first part of the lubrication height analysis accounts for the change in film thickness as a function of $S$ and material properties by assuming that the textures are flat.

Secondly, we consider the effect of fluid pressure on the deformation on the stripes. Here, we use a force balance at the fluid-solid interface to determine the compressive strain on the stripes. The force balance at the interface of the fluid and stripes states that:

$$p = \sigma \quad (S24)$$

where $p$ is the fluid pressure exerted on the stripe and $\sigma$ is the uniaxial compressive stress of the stripe. Since $p \sim \dfrac{U\eta a}{h^2}$ and $\sigma = E\varepsilon$, we expand equation S24 to include the strain and modulus and generate the following:

$$\dfrac{U\eta a}{\left(h_a + c \times \varepsilon\right)^2} = E\varepsilon. \quad (S25)$$

Equation S25 is the same as equation S15. Here, $\varepsilon$ is the strain to be determined. The solution for equation S25 shows that $\varepsilon$ is very small and results in a small deformation (see Deformation of striped PDMS surfaces section).

Finally, we observe that $\Delta h$ increases minimally at $S < S_c$ while significantly increasing at $S \geq S_c$ (Figure 2b). This last component is added to the previously described components of $h$ to generate the total film thickness. Because of the sudden change of $\Delta h$, the total film thicknesses $h$ and $H$ both show a critical increase at the transition point.

## Separation of the micro-EHL and macro-EHL lubrication forces

The critical transition of $h$ is used to separate EHL into two distinct regimes: micro-EHL and macro-EHL. The different fluid film thickness in two regimes results in two different physical models to predict the forces. Equations S19 and S20 are further simplified to predict the forces in micro-EHL regime. Because fluid dissipation is concentrated on stripes in this regime ($\frac{U\eta a}{h^2} A_a \gg \frac{U\eta b}{H^2} A_b \ \& \ \frac{U\eta}{h} A_a \gg \frac{U\eta}{H} A_b$), S16 & S17 become:

$$F_{N-micro} \sim \frac{U\eta a}{h^2} A_a \quad (S26)$$

$$F_{S-micro} \sim \frac{U\eta}{h} A_a \quad (S27)$$

where $h = h_a + c \times \varepsilon + \Delta h$ as indicated previously. The value of $h$ exhibits a steady increase in micro-EHL regime and it results in a constant $F_N$ and monotonic increase of $F_S$ (Figure 3a).

In the macro-EHL regime, the characteristic length changes from $a$ to $2R$, which results in a change of the scaling equations:

$$F_{N-macro} \sim \frac{U\eta(2R)}{H^2}(A_a + A_b) \quad (S28)$$

$$F_{S-macro} \sim \frac{U\eta}{H}(A_a + A_b) \quad (S29)$$

Equations S28 and S29 are obtained based on the assumption that textured surfaces are tribologically flat in the macro-EHL regime where $h \sim H \approx h_{smooth}$. The calculated normal force $F_N$ from equation S28 remains the same value as obtained from the micro-EHL calculations (Figure 3a). This predicted value of $F_N$ is in agreement with our experimental conditions.

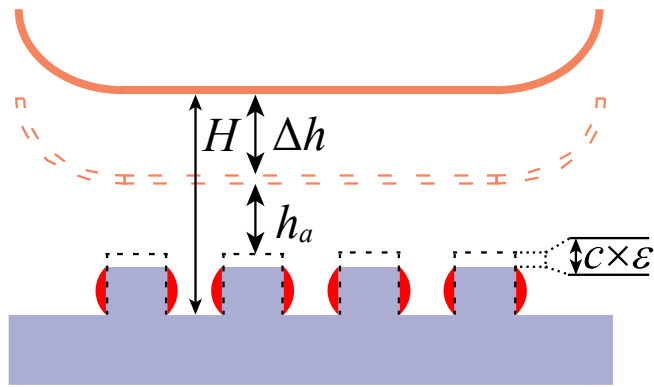

Figure S6 | Illustration of the three separate terms ($h_a$, $c \times \varepsilon$, and $\Delta h$) comprising the total film thickness $h$.

# Micro-EHL and macro-EHL prefactors in predicted $\mu$

In order to have a exact prediction of $\mu$, we use a prefactor $k$ in the scaling equations such that the friction coefficient takes the form of:

$$\mu_{lub} = k \frac{F_{S-prediction}}{F_{N-prediction}} = k\mu_{prediction} \quad (S30)$$

where $F_{S-prediction}$ comes from scaling equations S27 & S29 and $F_{N-prediction}$ comes from scaling equations S26 & S28. For PDMS-PDMS tribopairs, the prefactor $k$ in the micro-EHL regime is a function of the surface geometry ($k = f(geometry)$) as indicated by Figure S7a, which shows that $k$ is linearly correlated with $a$. However, in the macro-EHL regime, $k$ is not a function of $a$ as shown in Figure S7b. This is likely because in the macro-EHL regime, the two surfaces are far away from each other and surface geometry has limited effect on the shear and normal forces.

The striped surfaces made from mercaptoester and polyester are of fixed stripe width $a = 55$ μm and variable valley widths $b \in [35\mu m, 200\mu m]$ (Table S2). Since $k$ does not depend on $b$, the value of $k$ is exactly equal to 20 for PDMS-mercaptopher & PDMS-polyester tribopairs. As stated in the previous section, $E'$ for both kinds of tribopairs are the same. For PDMS-PEGDA/Alginate DN Hydrogel tribopair, $k = 9$. These observations suggest that $k$ is not only a function of surface geometry but also a function of material properties.

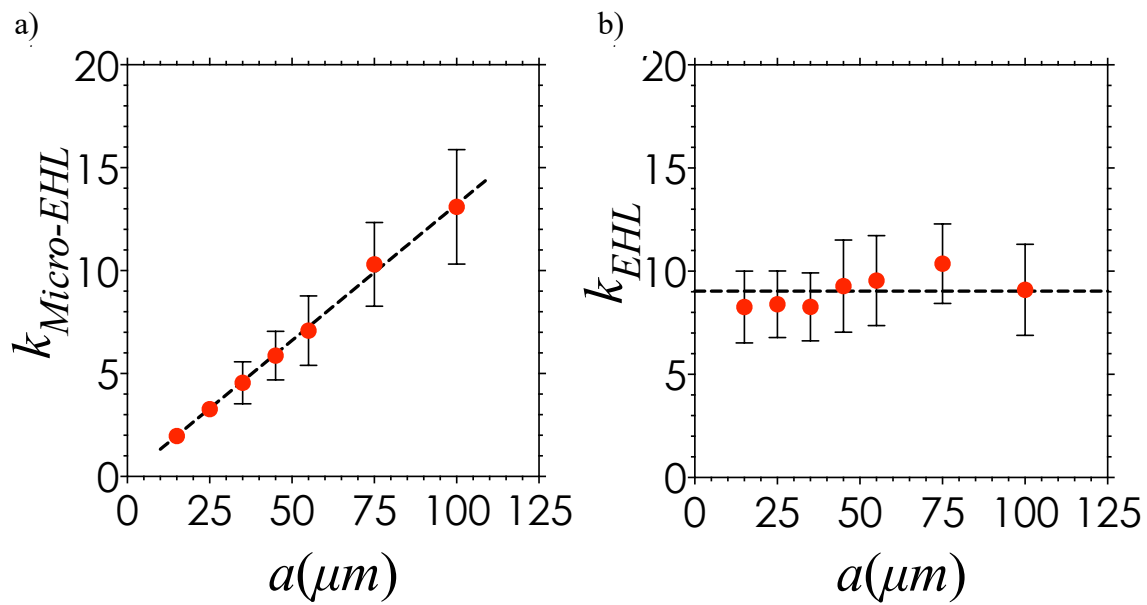

Figure S7 | Pre-factor *k* in micro-EHL (a) and macro-EHL regime (b) for PDMS-PDMS tribopair.

# Linear relation between friction, elasticity, and surface geometry

In Figure 4b, we obtained a linear correlation between the friction peak, normalized by the reduced elastic modulus, and the surface geometry. Here, we provide an explanation of the linear correlation based on dimensional analysis of the stripe length scales: (1) $\frac{a}{2R}$, which is the width of the stripe compared to the contact line, and (2) $\frac{a}{a+b}$, which is ratio of the width of the stripe to the length of a combined stripe-valley repeating unit.

In micro-EHL regime, we combine equation S26, S27 & S30 to obtain

$$\mu_{exp} \approx \mu_{lub} = k\mu_{prediction} = k\frac{U\eta a A_a / h^2}{U\eta A_a / h} = k\frac{h}{a}. \quad (S31)$$

Since Figure S7a indicates that $k \sim a$, for the friction peak of PDMS-PDMS tribopairs, equation S31 can be expressed as:

$$\mu_{c,exp} \sim h \quad (S32)$$

This relation for PDMS-PDMS tribopair holds, as seen in Figure S8a. Furthermore, our analysis in equation S23 combined with equation S32 can be extended as:

$$\mu_{c,exp} \sim h \approx h_{smooth}\sqrt{\frac{A_a}{A_{smooth}} \times \frac{a}{2R}} \approx \frac{h_{smooth}}{\sqrt{2R}}\sqrt{\frac{\frac{a}{a+b}A_{smooth}}{A_{smooth}} \times \frac{a}{1}} \approx h_{smooth}\sqrt{\frac{a}{2R}}\sqrt{\frac{a}{a+b}} \quad (S33)$$

Figure S8b indicates that $h_{smooth}$ is relatively constant for different surface geometries, and therefore equation S33 can be further simplified as:

$$\mu_{c,exp} \sim \sqrt{\frac{a}{2R}}\sqrt{\frac{a}{a+b}} \quad (S34)$$

Equation S34 indicates that the peak friction coefficient is directly proportional to the square root of the relative stripe dimensions $\frac{a}{2R}$ and $\frac{a}{a+b}$ as defined earlier. Finally, Figure S9 shows $R$ is also a constant regardless of the surface geometry. Therefore, equation S34 can be simplified to

$$\mu_{c,exp} \sim \frac{a}{\sqrt{a+b}} \quad (S35)$$

Numerous studies have shown that tribological phenomenon is directly related to the dimensionless friction force and normal force[9,13-17]: $\frac{F_S}{E'A}$ & $\frac{F_N}{E'A}$, where $A$ is the contact area. In our analysis, we define:

$$\widehat{F}_S = \frac{F_S}{E'A_a} \text{ and } \widehat{F}_N = \frac{F_N}{E'_{PDMS-PDMS}A_a} \quad (S35)$$

The reduced friction coefficient $\widehat{\mu}_c$ is then

$$\widehat{\mu}_c = \frac{\widehat{F}_S}{\widehat{F}_N} = \frac{\mu_c}{(E'/E'_{PDMS-PDMS})} \quad (S36)$$

The reduced friction coefficient $\widehat{\mu}_c$ shows a linear correlation with $a/(a+b)^{0.5}$ for all tribopairs with different materials as shown in Figure 4b. This linear correlation suggests that our model can be universally applied to variety of textured materials and can serve as a guideline to regulate friction on textured surfaces.

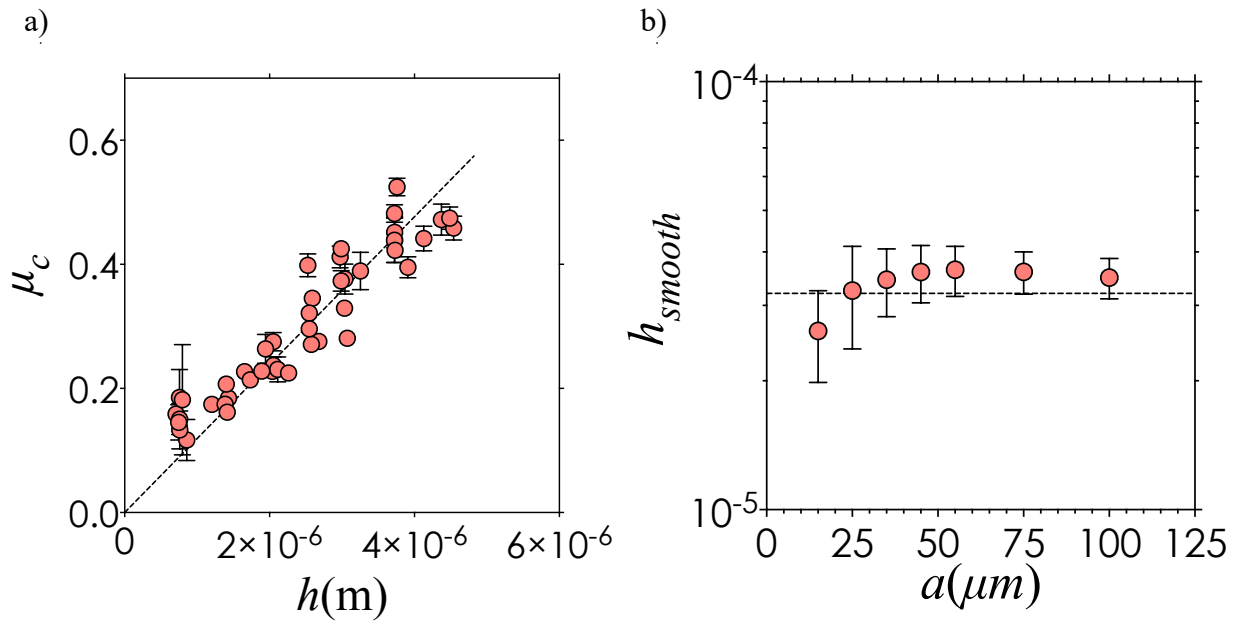

Figure S8 | (a) Experimental determination of the linear relation between $\mu_c$ and $h$ for PDMS-PDMS tribopair. (b) $h_{smooth}$ changes within a small range regardless of the surface geometry.

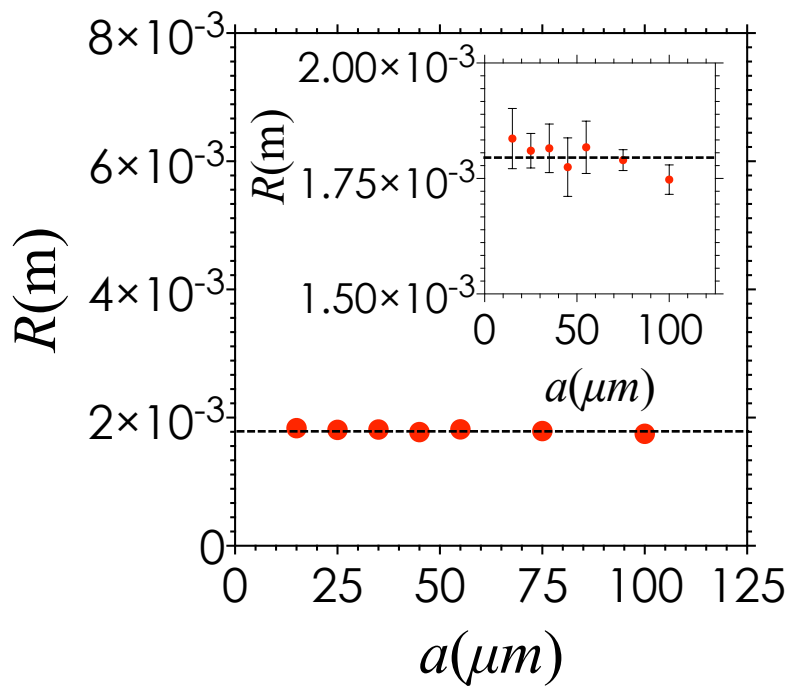

Figure S9 | Radius of contact area for striped surfaces with different $a$.

# Experimental data and predicted µ for all surface geometries

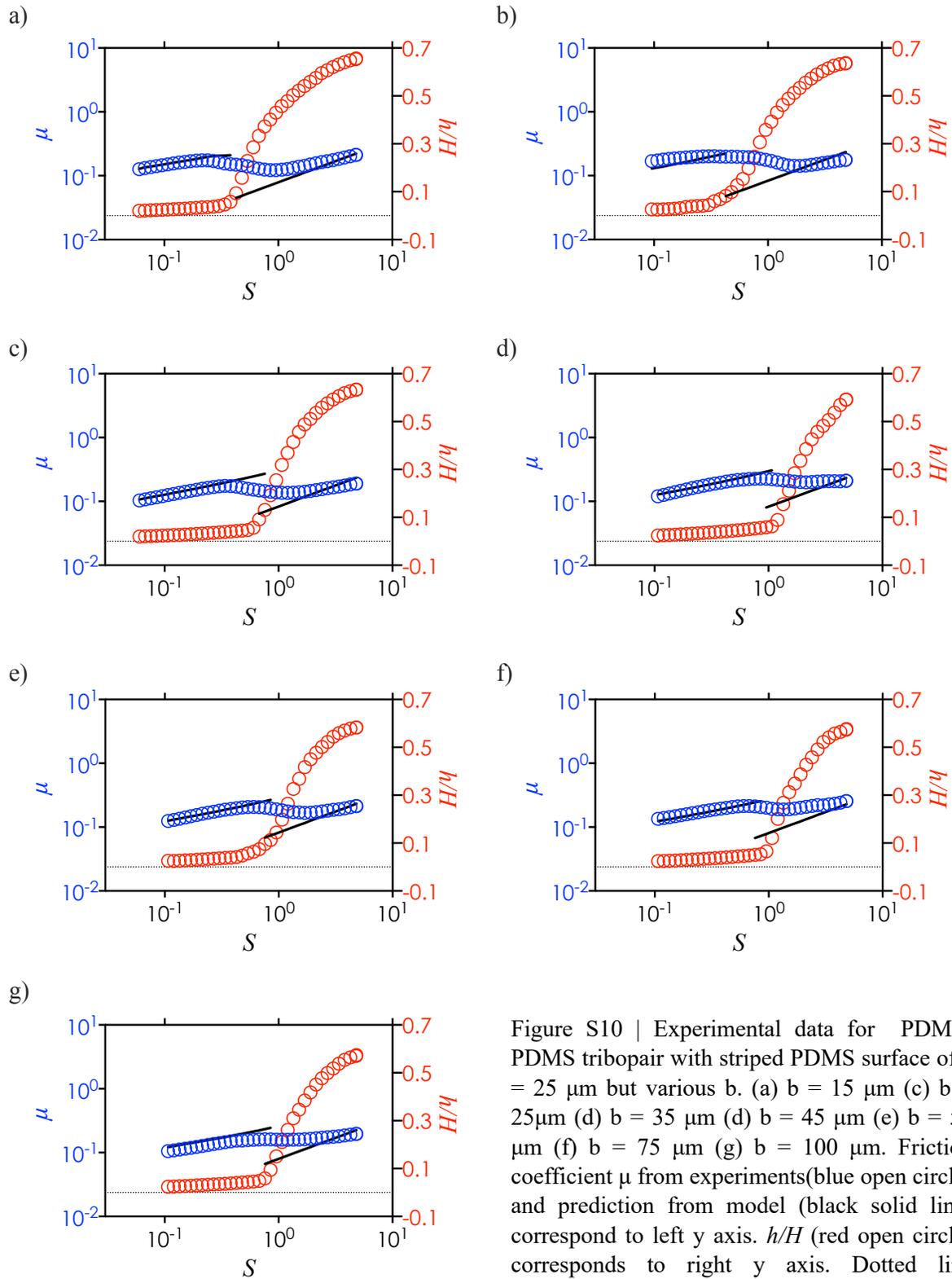

Figure S10 | Experimental data for PDMS-PDMS tribopair with striped PDMS surface of a = 25 µm but various b. (a) b = 15 µm (c) b = 25µm (d) b = 35 µm (d) b = 45 µm (e) b = 55 µm (f) b = 75 µm (g) b = 100 µm. Friction coefficient µ from experiments(blue open circle) and prediction from model (black solid line) correspond to left y axis. $h/H$ (red open circle) corresponds to right y axis. Dotted line represents $h/H = 0$.

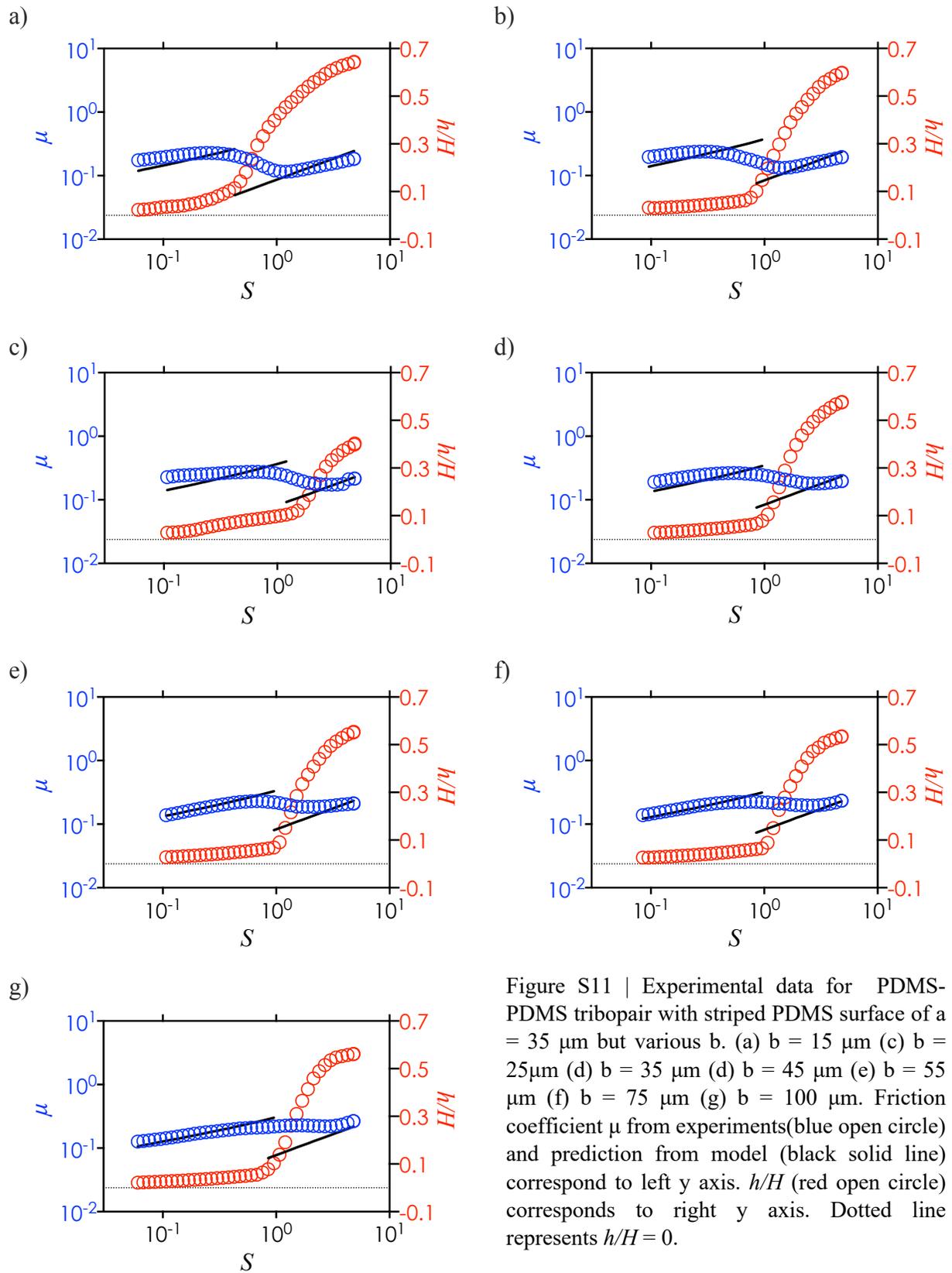

Figure S11 | Experimental data for PDMS-PDMS tribopair with striped PDMS surface of a = 35 μm but various b. (a) b = 15 μm (c) b = 25μm (d) b = 35 μm (d) b = 45 μm (e) b = 55 μm (f) b = 75 μm (g) b = 100 μm. Friction coefficient μ from experiments(blue open circle) and prediction from model (black solid line) correspond to left y axis. $h/H$ (red open circle) corresponds to right y axis. Dotted line represents $h/H = 0$.

S32

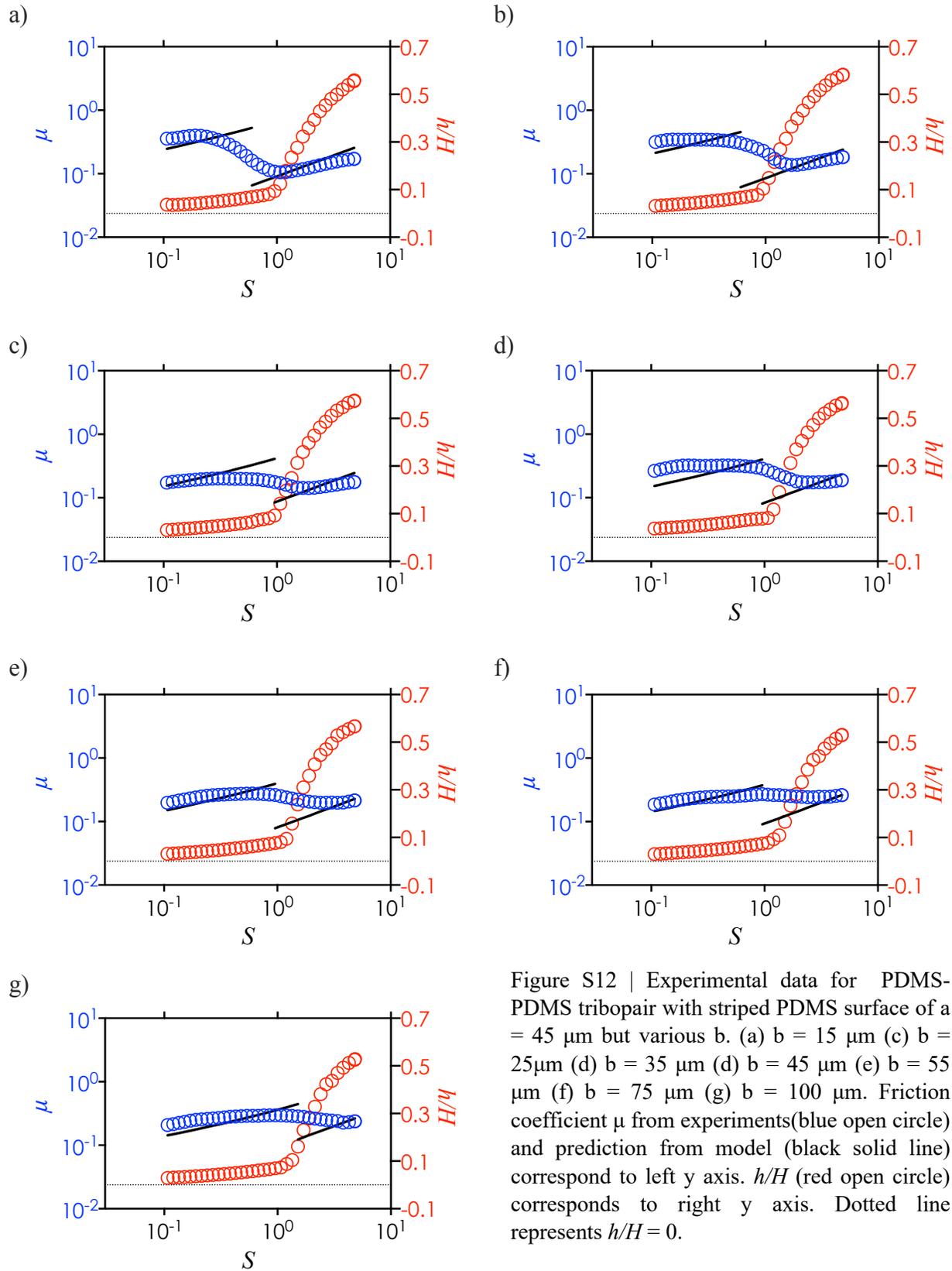

Figure S12 | Experimental data for PDMS-PDMS tribopair with striped PDMS surface of a = 45 μm but various b. (a) b = 15 μm (c) b = 25μm (d) b = 35 μm (d) b = 45 μm (e) b = 55 μm (f) b = 75 μm (g) b = 100 μm. Friction coefficient μ from experiments(blue open circle) and prediction from model (black solid line) correspond to left y axis. *h/H* (red open circle) corresponds to right y axis. Dotted line represents *h/H* = 0.

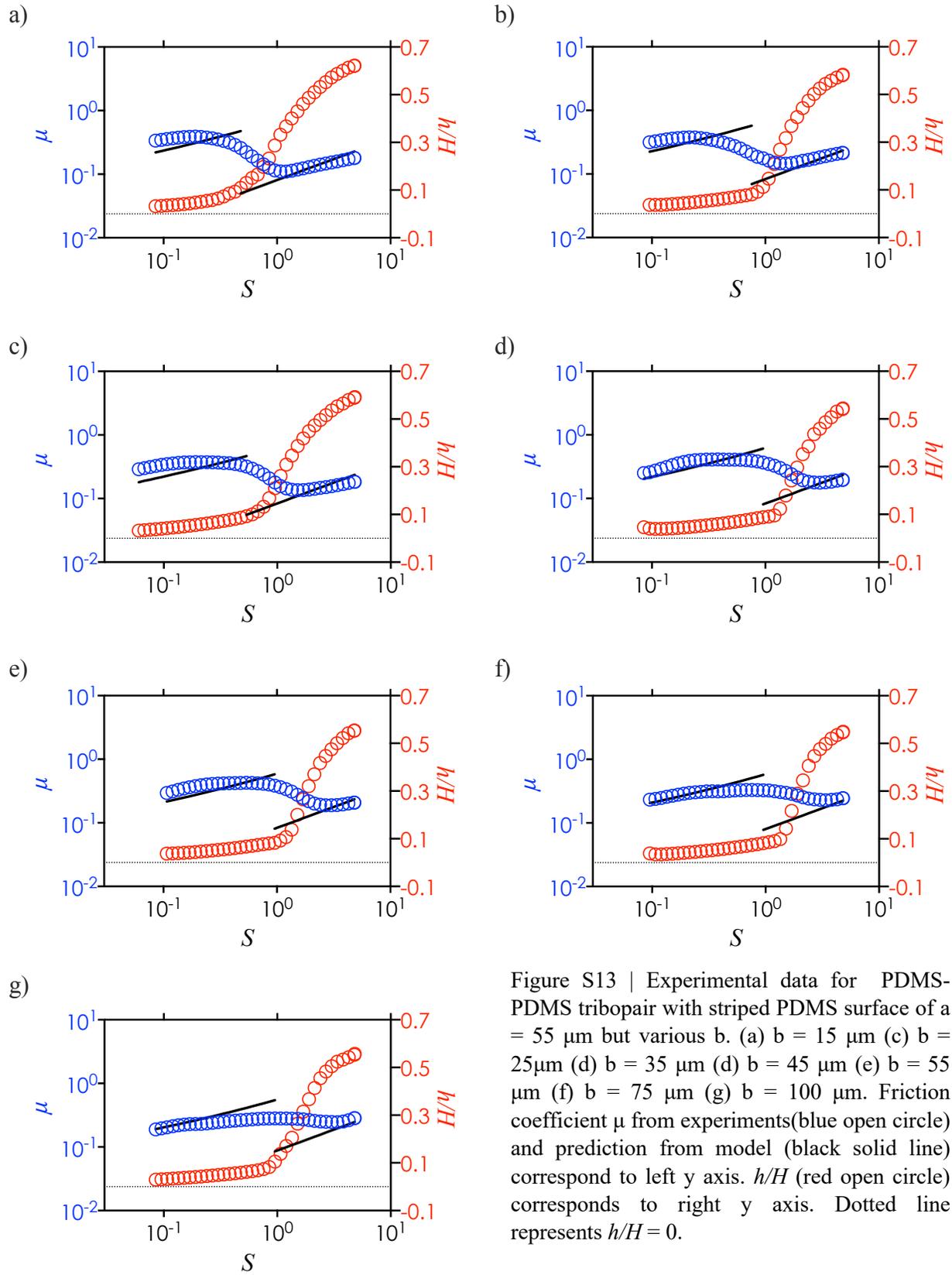

Figure S13 | Experimental data for PDMS-PDMS tribopair with striped PDMS surface of a = 55 μm but various b. (a) b = 15 μm (c) b = 25μm (d) b = 35 μm (d) b = 45 μm (e) b = 55 μm (f) b = 75 μm (g) b = 100 μm. Friction coefficient μ from experiments(blue open circle) and prediction from model (black solid line) correspond to left y axis. $h/H$ (red open circle) corresponds to right y axis. Dotted line represents $h/H = 0$.

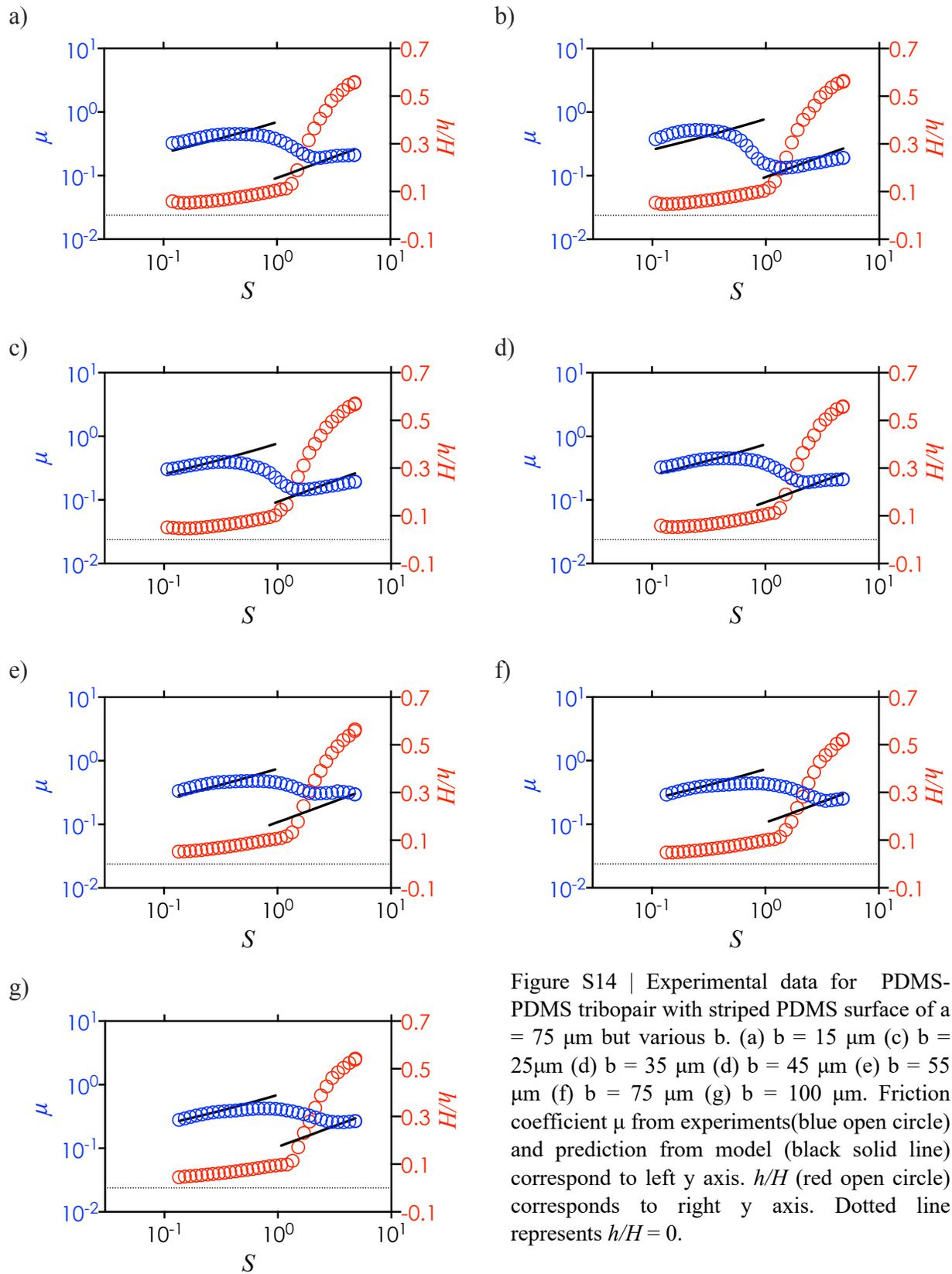

Figure S14 | Experimental data for PDMS-PDMS tribopair with striped PDMS surface of a = 75 μm but various b. (a) b = 15 μm (c) b = 25μm (d) b = 35 μm (d) b = 45 μm (e) b = 55 μm (f) b = 75 μm (g) b = 100 μm. Friction coefficient μ from experiments(blue open circle) and prediction from model (black solid line) correspond to left y axis. $h/H$ (red open circle) corresponds to right y axis. Dotted line represents $h/H = 0$.

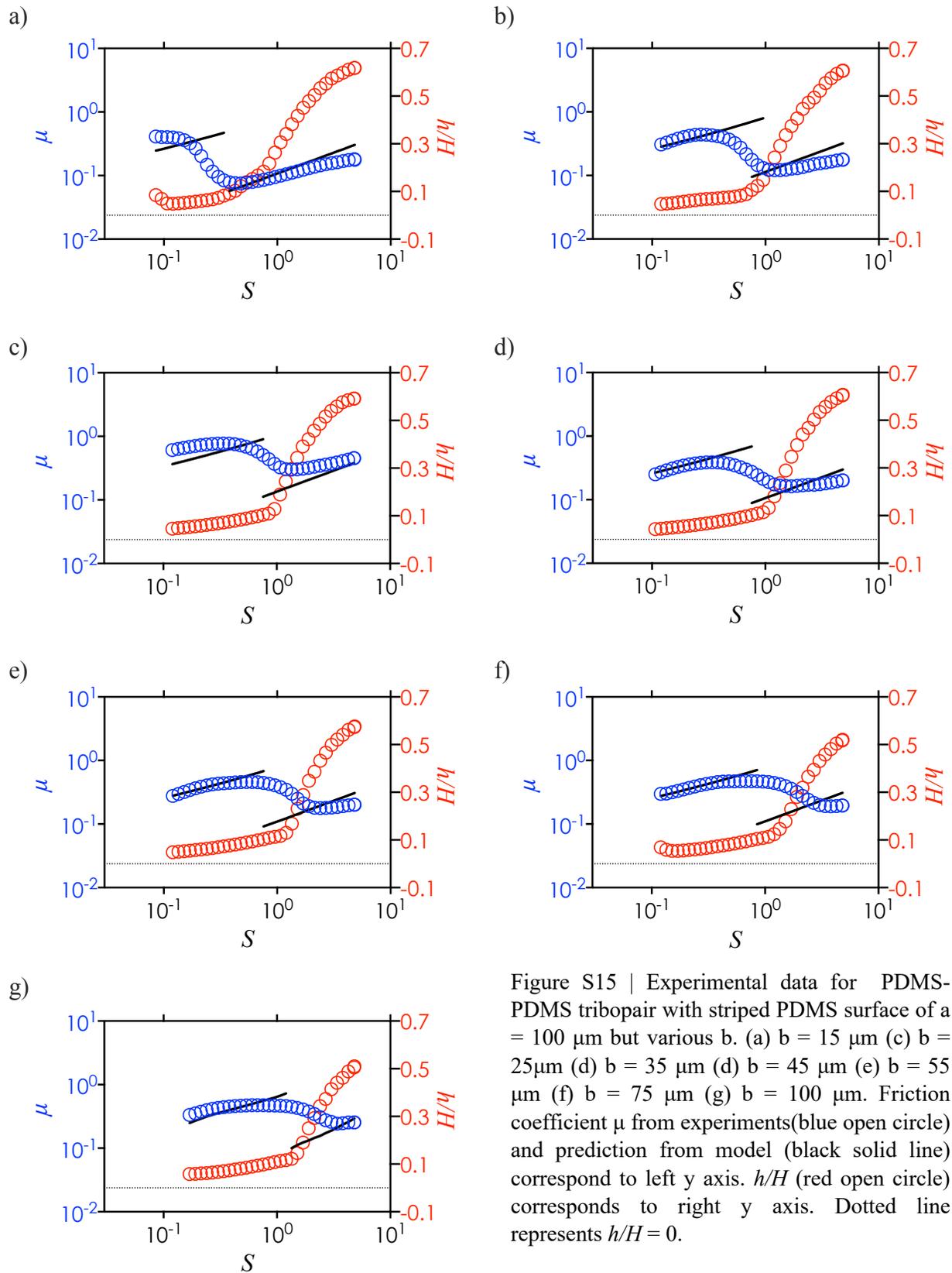

Figure S15 | Experimental data for PDMS-PDMS tribopair with striped PDMS surface of a = 100 μm but various b. (a) b = 15 μm (c) b = 25μm (d) b = 35 μm (d) b = 45 μm (e) b = 55 μm (f) b = 75 μm (g) b = 100 μm. Friction coefficient μ from experiments(blue open circle) and prediction from model (black solid line) correspond to left y axis. $h/H$ (red open circle) corresponds to right y axis. Dotted line represents $h/H = 0$.

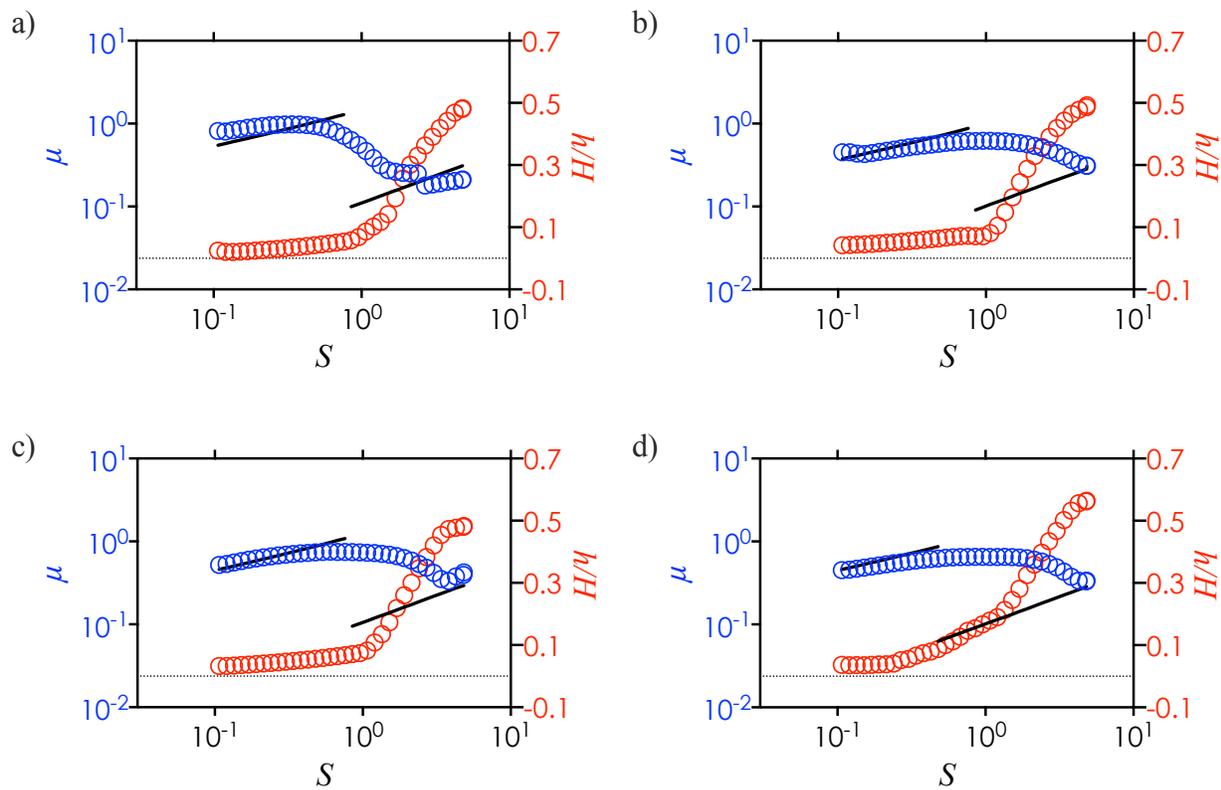

Figure S16 | Experimental data for PDMS-mercaptoester tribopair with striped mercaptoester surface of a = 55 μm but various b. (a) b = 35 μm (c) b = 75μm (d) b = 100 μm (d) b = 200 μm. Friction coefficient μ from experiments(blue open circle) and prediction from model (black solid line) correspond to left y axis. $h/H$ (red open circle) corresponds to right y axis. Dotted line represents $h/H = 0$.

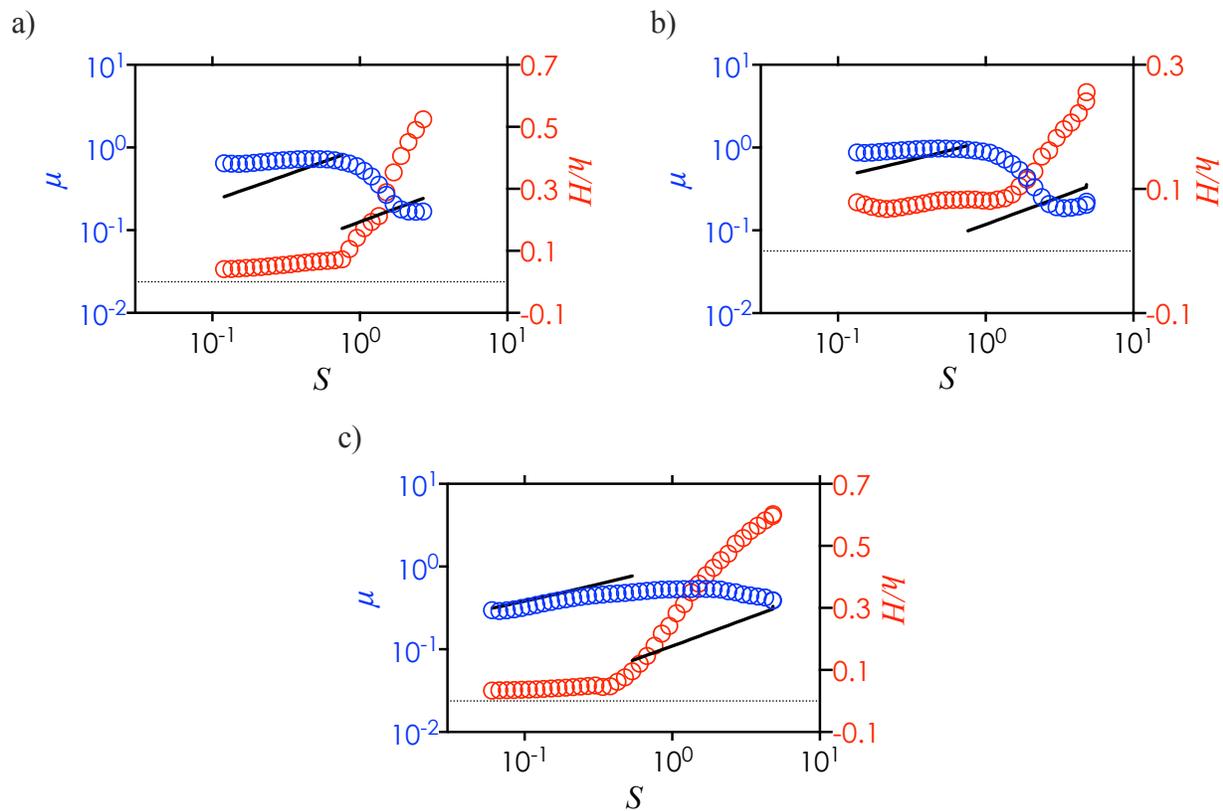

Figure S17 | Experimental data for PDMS-polyester tribopair with striped polyester surface of a = 55 μm but various b. (a) b = 35 μm (b) b = 75μm (c) b = 100 μm. Friction coefficient μ from experiments(blue open circle) and prediction from model (black solid line) correspond to left y axis. *h/H* (red open circle) corresponds to right y axis. Dotted line represents *h/H* = 0.

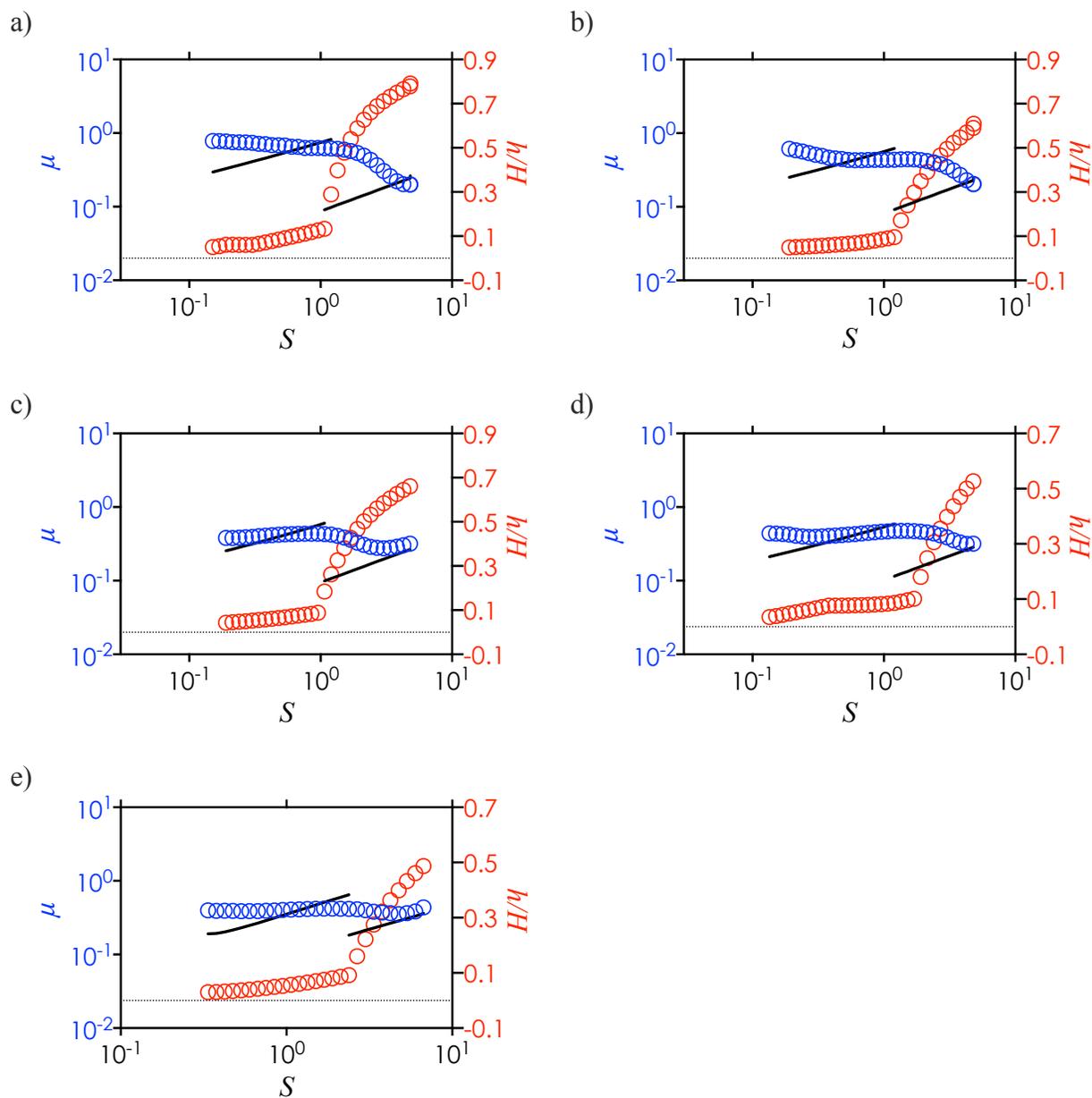

Figure S18 | Experimental data for PDMS-DN Hydrogel tribopair with striped hydrogel surface of a = 55 μm but various b. (a) b = 35 μm (b) b = 55μm (c) b = 75μm (d) b = 100 μm (e) b = 200μm. Friction coefficient μ from experiments(blue open circle) and prediction from model (black solid line) correspond to left y axis. $h/H$ (red open circle) corresponds to right y axis. Dotted line represents $h/H = 0$.